\documentclass[aip,pop,reprint]{revtex4-1}

\usepackage{graphicx}
\usepackage{dcolumn}
\usepackage{bm}
\usepackage{amssymb}
\usepackage{amsmath}
\usepackage{siunitx}

\begin{document}


\title{Ion Temperature Evolution in an Ultracold Neutral Plasma} 



\author{P. McQuillen}
\email[]{patrickmcquillen@rice.edu}
\author{T. Strickler}
\author{T. Langin}
\author{T. C. Killian}
\affiliation{Rice University, Department of Physics and Astronomy, Houston, Texas 77005}


\date{\today}

\begin{abstract}
We study the long-time evolution of the ion temperature in an expanding ultracold neutral plasma using spatially resolved, laser-induced-fluorescence spectroscopy. Adiabatic cooling reduces the ion temperature by an order of magnitude during the plasma expansion, to temperatures as low as \SI{0.2}{\kelvin}. Cooling is limited by heat exchange between ions and the much hotter electrons. We also present evidence for an additional heating mechanism and discuss possible sources.  Data are described by a model of the plasma evolution, including the effects of ion-electron heat exchange. We show that for appropriate initial conditions, the degree of Coulomb coupling of ions in the plasma increases during expansion.
\end{abstract}


\maketitle 

\section{Introduction}

Ultracold neutral plasmas (UNPs), formed by photoionizing laser-cooled atoms near threshold, have ion temperatures around \SI{1}{\kelvin} and tunable electron temperatures from \SIrange{1}{1000}{\kelvin}. \cite{kpp07} One of the most important topics studied in UNPs is the physics of strongly coupled plasmas, which have a ratio of nearest neighbor potential energy to kinetic energy larger than one. This is quantified by the Coulomb coupling parameter
\begin{equation}
\label{eq:GammaCouplingPlasma}
\Gamma=\frac{e^2}{4\pi \varepsilon_0 a_{ws}}/\left(k_B T\right),
\end{equation}
where  $a_{ws}\equiv\left(4\pi n/3\right)^{-1/3}$ is the average interparticle spacing (i.e. Wigner-Seitz radius) and $T$ is the temperature of the particle species of interest. Ions in UNPs have $\Gamma_{i}\approx 2-4$, and they have been used to study the equilibration of strongly coupled Coulomb systems after a potential quench \cite{mur01,kon02prl,mck02,csl04,cdd05,bdl11,lbm13} and collision rates beyond the regime of validity of Landau-Spitzer theory. \cite{bcm12}  Strong coupling is of interest in many different plasma environments, \cite{ich82} and the equilibration dynamics observed in UNPs in particular has been connected to laser-produced plasmas formed from clusters and solids \cite{mur07pop} and proposed as a possible limitation in the brightness of photoemitted electron beams. \cite{mbw13}

A central focus of research on UNPs is the creation of more strongly coupled plasmas to study  collisions, transport, collective modes, and many other phenomena across a wider variety of coupling strengths into the liquid regime. Proposals to increase $\Gamma_i$ by modifying the plasma after its creation are sensitive to the long-term evolution of the ion temperature. Here, we report measurement and modelling of the ion-temperature evolution of expanding UNPs, which has never been experimentally studied before because of the difficulty in distinguishing the small thermal motions of ions from  the large, hydrodynamic ion expansion velocity. \cite{kkb00,rha02,lgs07} Our results show significant ion adiabatic cooling, with ion temperatures decreasing up to an order of magnitude.  We also observe significant contributions to the ion temperature from ion-electron thermalization, as well as an additional source of ion heating we ascribe to deviations of the plasma density from an ideal spherical Gaussian. 

This study adds to prior studies of fundamental plasma properties conducted using UNPs, such as the plasma creation process, \cite{kkb00,mck02,rha02,kon02prl} collective modes of ions \cite{cmk10,mcs13} and electrons, \cite{kkb00,tro12,lpr12} and formation and ionization of Rydberg atoms in the plasma. \cite{klk01,bro08,rtn00}

There is long-standing interest in the problem of a plasma expanding into vacuum, which typically dominates the dynamics of plasmas created with pulsed lasers, \cite{pmo94} such as in experiments pursuing inertial confinement fusion, \cite{lin95} x-ray lasers, \cite{dai02} or the production of energetic ($>$MeV) ions through irradiation of solids, \cite{ckz00,skh00} thin foils, \cite{mgf00} and clusters. \cite{shh07} 

For ultracold neutral plasmas, an early hydrodynamic model and numerical study, which included the effects of three-body recombination and other inelastic processes, \cite{rha03} accurately described experimental observations of the ion density evolution. \cite{kkb00} An adiabatic, self-similar solution of the Vlasov equations \cite{bku98,dse98,kbt02,kby03,lgs07} describes the expansion dynamics well across a wide range of initial conditions, and this is the starting point for understanding the experiments described here. This model predicts adiabatic cooling of electrons, which has been confirmed through the expansion dynamics, \cite{lgs07,gls07} electron loss, \cite{rfl04} and the TBR rate. \cite{fzr07} The model predicts adiabatic cooling of the ions as well, and it was pointed out that this should result in an increase of correlations to $\Gamma_i \approx 10$  before the density and collision rate drop so much that correlations freeze out. \cite{ppr04,ppr05PRL} Small effects of ion correlations  on the expansion have been discussed in conjunction with a comprehensive model. \cite{ppr04} Our work provides the first test of these predictions of the evolution of the ion temperature.

Adiabatic cooling has been observed in trapped non-neutral plasmas when reducing confinement for both pure ion \cite{lpt91} and antiproton plasmas. \cite{gkm11,mh12} It is important in the interstellar medium and solar wind \cite{swk09,mhl12,cmg13} and it has been discussed for ions in a plasma created by laser irradiation of a solid. \cite{bcc05,cbc06}

\section{Experimental Methods}
UNPs are created by photoionizing laser-cooled  $^{88}\mathrm{Sr}$ atoms in a magneto-optical trap. \cite{kpp07} This results in a spherical Gaussian density distribution (initial $e^{-1/2}$ density radius $\equiv \sigma_0 \sim$ \SIrange{1}{2}{\milli\metre}) of a few hundred million ions (and electrons) with a peak density of up to \SI{5e16}{\per\metre\cubed}.

Spatially and temporally resolved laser induced fluorescence spectroscopy on the ion's principal transition at \SI{422}{\nano\metre} permits {\it{in situ}} probing of the local kinetic energy and density of the ions. \cite{lgs07,cgk08}
We excite fluorescence in a  \SI{1}{\milli\metre} thick sheet  that bisects the plasma.
 Fluorescence emitted perpendicular to the  sheet is collected by a 1:1 optical relay and 4$\times$ objective/ocular magnification stage that images onto an intensified charge coupled device (ICCD), with a pixel size of \SI{13}{\micro\metre}. This allows regional analysis of small volumes of the plasma with roughly constant density and bulk expansion velocity. \cite{lcg06,cgk08} The total system resolution is mainly limited by the ICCD, which suffers from blooming during the amplification process. To minimize this effect, we experimentally obtained the point spread function of the ICCD \cite{pcmThesis12} and deconvolve each image in post analysis. \cite{cas07}

Creation and probing of the plasma is repeated with a 10\,Hz repetition rate, and the LIF-laser frequency is scanned to build up an excitation spectrum. The spectrum of the region of interest is fit to a Voigt profile. We assume a Lorentzian-component width dominated by the laser and natural linewidths (\SI{6}{\mega\hertz}, \SI{20}{\mega\hertz}, respectively) and the Doppler-broadened Gaussian width is a fit parameter related to a local ion temperature. \cite{cgk08}
 The signal is proportional to density and can be calibrated using absorption imaging, \cite{scg04} with a resulting uncertainty in density of approximately $\pm $ \SI{20}{\percent}.

To minimize the contamination of ion temperature measurements by expansion velocity, \cite{cgk08} temperatures are measured for regions that are only one pixel wide along the LIF beam axis and \SI{1}{\milli\metre} perpendicular to it. To improve statistics, we analyze 79 regions covering a 1 mm$^2$ area, and compute an average temperature from the measurements. 
Error bars on temperature measurements are statistical uncertainty in the resulting mean. The averaging over multiple regions introduces a maximum relative density variation of $e^{-1/8}= 0.88$ from the center region to the outermost region, for $\sigma_0 = $\SI{1}{\milli\metre}. All densities we quote are an average of the regional densities, $\geq$ \SI{95}{\percent} of the peak density.

\section{Overview of ultracold plasma evolution time scales}
\begin{figure}[h]
\includegraphics{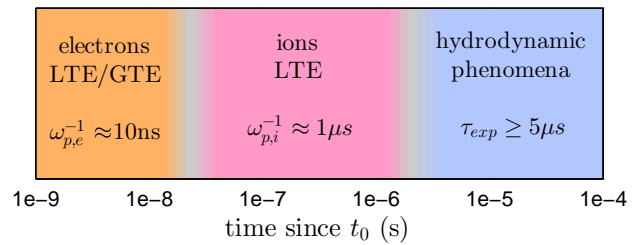}
\caption{\label{fig:TimeScales} Three main stages of UNP evolution. Electrons  reach global thermal equilibrium (GTE) on a timescale approximately equal to the  inverse electron plasma frequency, $\omega_{pe}^{-1} = \sqrt{\varepsilon_0 m_e/n_e e^2}$.
Ions come to local thermal equilibrium (LTE) on a timescale of the  inverse ion plasma frequency. Hydrodynamic effects such as plasma expansion occur on the longest timescale.}
\end{figure}

The plasma is unconfined and undergoes a complex evolution as it expands into the surrounding vacuum, \cite{kpp07} with dynamics that can be divided into three distinct time scales as shown in Figure \ref{fig:TimeScales}.

Global thermal equilibrium for electrons is established on a very fast timescale compared to the physics of interest in this paper at a temperature determined by  \cite{rha03,kpp07}
\begin{equation}
\frac{3 k_B T_e}{2}=\hbar\omega_l- E_{i}.
\end{equation}
Here, $\hbar\omega_l$ is the combined energy of the photoionizing photons and $E_{i}$ is the ionization energy of a ground state atom. We can create UNPs with initial electron temperatures of \SIrange{1}{1000}{\kelvin}, however we typically stay above \SI{40}{\kelvin} to avoid three body recombination effects, which become important when $\Gamma_e \geq 0.1$. \cite{gls07} We ionize a maximum of about 25\% of the atoms for $T_e\lesssim 150$\,K, but this approaches 100\% near an auto-ionizing resonance (Te $\approx$ \SI{430}{\kelvin}). \cite{gc68} All but a few percent of the electrons are trapped by the space-charge field of the ions. \cite{kkb99}

Next, the ions come to local thermal equilibrium on a time scale comparable to the inverse of their fundamental oscillation frequency, $\omega_{pi}^{-1} = \sqrt{\varepsilon_0 m_i/n_i e^2} \sim$ \SI{1}{\micro\second}. Initial ion  velocities and positions are inherited from the laser-cooled atoms,  resulting in low kinetic energy and high potential energy because the spatial distribution is completely random. Coulomb energy is converted into ion kinetic energy in a process called disorder-induced heating (DIH) or correlation heating, \cite{mur01,scg04} which yields a density-dependent, equilibrium ion temperature (T$_{DIH}\propto n^{-1/3} \sim$ \SI{1}{\kelvin}). Disorder-induced heating limits the ion's Coulomb coupling parameter to $2 \lesssim \Gamma_i \lesssim 4$, with the variation determined by electron screening. \cite{lbm13} 
Strictly speaking ions never achieve global thermal equilibrium in our experiments because of the long timescale for thermal transport. \cite{dha04}


In the next stage of evolution, the plasma expands into the surrounding vacuum in a hydrodynamic fashion, driven by the electron thermal pressure.
For a quasi-neutral, perfect Gaussian density distribution, assuming global thermal equilibrium of ions and electrons and negligible  inelastic collision processes and electron-ion thermalization, the expansion is described by a self-similar expansion 
\begin{eqnarray}
\sigma^2(t) &=& \sigma_0^2(1+t^2/\tau_{exp}^2), \label{eq:sizeevolution}\\
T_{i,e}(t) &=& T_{i,e}(0)/(1+t^2/\tau_{exp}^2), \label{eq:tempevolution}
\end{eqnarray}
where the expansion timescale,
$\tau_{exp}=\sqrt{{m_i \sigma_0^2}/{k_B [T_e(0)+T_i(0)]}}\approx 10\,\mu\mathrm{s}$, is set by initial size and temperatures after electrons and ions reach global and local equilibrium respectively. \cite{dse98,lgs07} The electron temperature dominates, which allows one to indirectly measure $T_e(0)$ by fitting the evolution of the cloud size. Equations \ref{eq:sizeevolution} and  \ref{eq:tempevolution} also show that $T_{e,i}\sigma^2=constant$, reflecting the adiabatic cooling of both species as the cloud expands. We will discuss below how these equations must be modified to account for heat exchange between electrons and ions, but they provide good intuition for plasma dynamics.

\begin{figure}[h]
\includegraphics{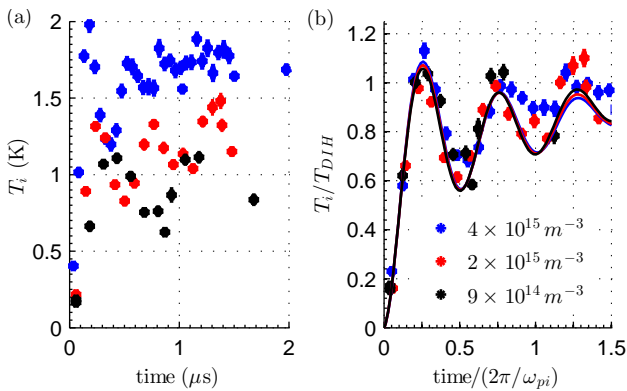}
\caption{\label{fig:KEOs} Early time measurements of ion kinetic energy widths for various density UNPs with $T_e =$ \SI{430}{\kelvin} (a) Disorder induced heating temperatures vary with density as shown. (b) Scaling temperature by the predicted DIH magnitude and time by the ion plasma oscillation  period results in close to universal behavior. The solid lines are results from molecular dynamics simulation of an equilibrating plasma which are fit to experimental data to determine ion density.}
\end{figure}

\section{Initial Ion Equilibration: Disorder-Induced Heating, Kinetic-Energy Oscillations,  and Density Calibration} \label{Sec:DIHandKEO}
The initial equilibration of the ions has been extensively discussed previously, \cite{mur01,kon02prl,mck02,csl04,cdd05,mur07pop,bdl11,lbm13} but we describe it here in order to give a complete picture of the temperature evolution and because early-time heating dynamics provide the most accurate determination of the plasma density.

Figure \ref{fig:KEOs} shows the increase of ion kinetic energy due to initial disorder (DIH) and the subsequent kinetic energy oscillations (KEOs) at close to twice the ion plasma frequency $\omega_{pi}$ that result from the evolution of spatial correlations in the equilibrating strongly coupled plasma.  The equilibrium temperature from DIH for a homogeneous plasma, is set by density ($\propto n^{1/3}$) and electron temperature (via electron-screening). It can be calculated from  \cite{mur01}
\begin{equation}
T_{DIH}\equiv\frac{2}{3k_B} \frac{e^2}{4\pi\varepsilon_0 a_{ws}}\left|\tilde{U}+\frac{\kappa}{2}\right|
\end{equation}
using tabulated molecular dynamics simulation results \cite{hfd97} for $\tilde{U}$, the excess particle energy in units of $e^2/4\pi \varepsilon_0 a_{ws}$.  $\kappa \equiv a_{ws}/\lambda_D$ is the screening parameter for electron Debye length $\lambda_D=(\varepsilon_0 k_B T_e/n_{e} e^2)^{1/2}$. A closed-form description of KEO dynamics does not exist, and it can only be predicted with molecular dynamics simulations.
The oscillation frequency has been shown by particle-in-cell Yukawa simulations, \cite{bdl11} tree-code algorithms \cite{mbw13} and full molecular dynamics simulations \cite{glh10} to agree well with $\omega_{pi}$, although electron screening softens the ion-ion interaction and slows the oscillations. \cite{lb11,bdl11} This deviation is small but observable  for our conditions. As seen in Fig. \ref{fig:KEOs}(b), when time is scaled by $2\pi/\omega_{pi}$, and ion temperature is scaled by $T_{DIH}$, the various plasma exhibit similar behavior, and oscillations damp to a scaled temperature close to unity.

All presented temperatures are fit parameters assuming a Voigt spectral profile with Maxwellian velocity distribution.
It has been shown,  \cite{bdl11} however, that the velocity distribution shows a non-Maxwellian, high-velocity tail that relaxes on the same timescale as the oscillations, so Voigt fits will underestimate the ion RMS kinetic energy until the ions have achieved local thermal equilibrium. Therefore, to interpret  our early-time measurements we compare them to a library of ion velocity distributions from molecular dynamics simulations of a homogeneous plasma. \cite{pohlPrv15} We use the numerical data to produce simulated LIF spectra that are fit with the same procedure as experimental data (Fig. \ref{fig:KEOs} (b)). Assuming an electron temperature given by the photoionizing-laser photon energy we can allow density to be a fit parameter, and the fit is highly constrained because the KEO frequency and the temperature both depend on the density. Good confidence in the measured density and $T_{DIH}$ is important for accurate simulation of the full ion temperature evolution as discussed below. Note that the measured temperature only approaches $T_{DIH}$ as the oscillations damp.

Fits of experimental data to numerical results are presented as insets throughout the paper. A full study of this density-fitting method will be given elsewhere. \cite{tbp15} It is more accurate than using LIF intensity calibrated with absorption imaging, \cite{cgk08} but the two methods agree well, yielding an absolute uncertainty in density on order of $\pm 10 $\%.

\section{Ion Adiabatic Cooling}
\begin{figure}[h]
\includegraphics[width=3.25in]{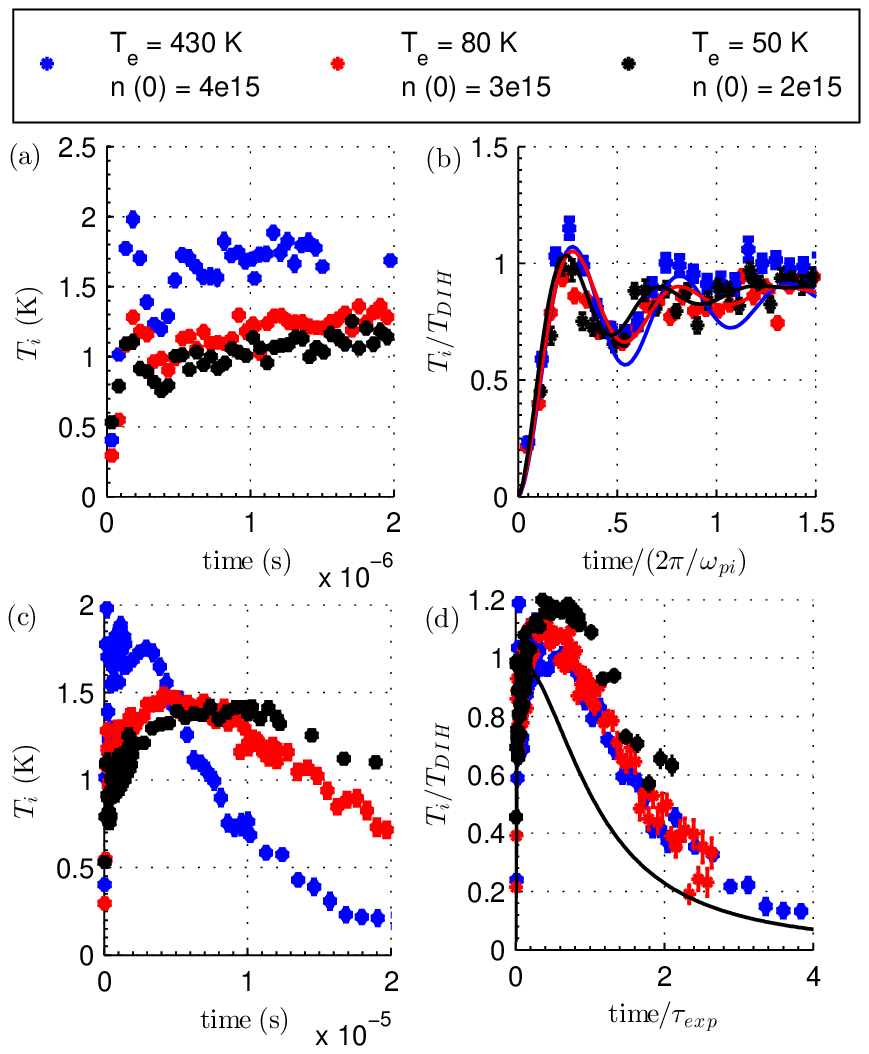}
\caption{\label{fig:Expansion} Ion temperature evolution for UNPs with similar initial plasma size ($\sigma_0 = $ \SI{1}{\milli\metre}) and comparable ion densities ($n_i(0) = $ \SIrange{2e15}{4e15}{\per\metre\cubed}) and  different initial electron temperatures given in the legend. (a) Lower initial electron temperature produces stronger electron shielding and lower temperature after DIH. (b) Scaling the time and temperature axes yields universal curves that only vary with electron screening. The comparison of experimental data with molecular dynamics simulations (solid lines) is used to determine the local ion density. (c) Lower electron temperature yields slower expansion and ion cooling. (d) Scaling the temperature by $T_{DIH}$ and time by the expansion time $\tau_{exp}$ approximately collapses the data onto a universal curve, confirming that adiabatic cooling  dominates the long-time evolution. The line is the prediction of the ion temperature model (Eqs.\ \ref{eq:TempEvolModela}- \ref{eq:CorrelationEnergyEvolution}) including correlation effects and cloud expansion only, \cite{ppr04} omitting electron-ion thermalization}
\end{figure}

\begin{figure}[h]
\includegraphics[width=3.25in,]{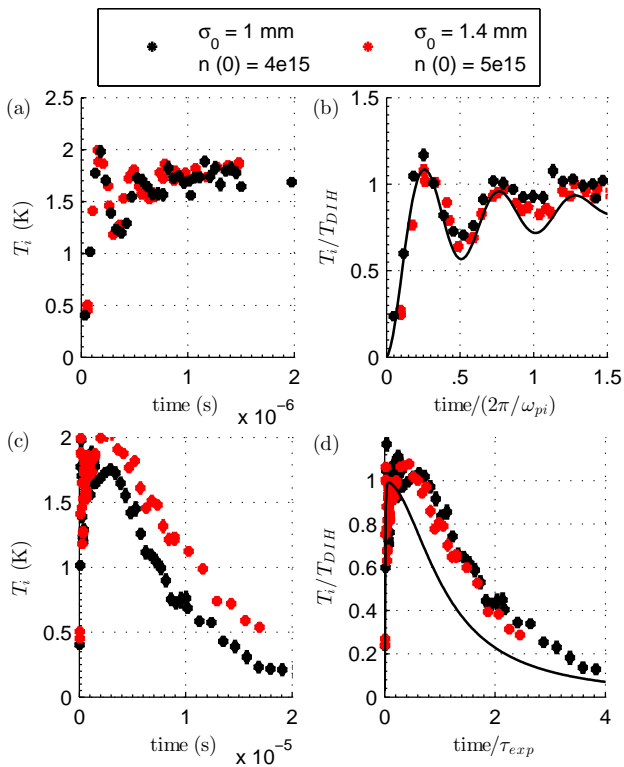}
\caption{\label{fig:ExpansionSizes} Long time ion temperature evolution for UNPs with similar initial electron temperature ($T_e(0) =$ \SI{430}{\kelvin}) and comparable ion densities yet different initial plasma sizes (details in the legend). (a) Plasma size does not effect local DIH dynamics. (b) Comparison with molecular dynamics simulations is used to determine local ion density. (c) Larger initial cloud size yields slower expansion and ion cooling. (d) Scaling the temperature by $T_{DIH}$ and time by the expansion time $\tau_{exp}$ approximately collapses the data onto a universal curve.}
\end{figure}

On a hydrodynamic timescale of $\sim $ \SI{10}{\micro\second} the plasma cloud will expand due to the thermal pressure of the electrons, leading to adiabatic cooling of electrons and ions. When ion temperatures are measured for longer evolution times we clearly see cooling of the ions by up to an order of magnitude, as demonstrated in Figs. \ref{fig:Expansion}  and \ref{fig:ExpansionSizes}.

To establish that we are observing adiabatic cooling of the ions, we check the scaling predicted by Eq.\ \ref{eq:tempevolution}. According to this simple model, the only two parameters determining the ion temperature are the initial temperature, set by DIH, and the expansion time $\tau_{exp}$. Figure \ref{fig:Expansion} (b) shows the evolution of the ion temperature scaled by those two parameters for various initial electron temperatures holding plasma size and  initial peak density fixed.
Figure \ref{fig:ExpansionSizes} (b) shows a similar study varying the plasma size for similar initial electron temperature
and ion density.
In both cases, the scaled data approximately collapse onto  universal curves, showing that adiabatic cooling, described by Eq.\ \ref{eq:tempevolution}, dominates the ion temperature evolution. We observe temperatures as low as 0.2 K, which are the lowest temperatures ever measured for ions in local thermal equilibrium in an ultracold neutral plasma.

%

\section{Modelling UNP Dynamics}
\label{sec:Dynamics}
To more accurately model the data, we use a hybrid approach first developed in Ref.\ \cite{ppr04PRA} that combines a hydrodynamic treatment of the evolution of the plasma size and electron and ion temperatures with terms derived from a kinetic and numerical treatment of the effects of correlations. The model takes the form of differential equations describing the evolution of the Gaussian size parameter $\sigma$, the electron and ion temperatures, and the expansion parameter $\gamma$. The hydrodynamic expansion velocity for the ions at position $\bf{r}$ with respect to plasma center, is $\bf{v}(\bf{r},t)=\gamma(t)\bf{r}$. The differential form of the equations allows inclusion of new terms to treat electron-ion thermalization.

\begin{align}
\frac{\partial\sigma^2(t)}{\partial t}&=2\gamma(t)\sigma^2(t) \label{eq:TempEvolModela} \\
\frac{\partial\gamma(t)}{\partial t}&=\frac{k_B T_e(t)+U_{ii}(t)/3}{m_i\sigma^2(t)}-\gamma^2(t) \label{eq:TempEvolModelb}\\
\begin{split}
\frac{\partial T_i(t)}{\partial t}&=-2\gamma(t)T_i(t)-\frac{2}{3}\left(\gamma(t)U_{ii}(t)+\frac{\partial U_{ii}(t)}{\partial t}\right)+\\
&2\frac{m_e}{m_i}{\gamma_{ei}}(t)T_e(t) \label{eq:TempEvolModelc}
\end{split}\\
\frac{\partial T_e(t)}{\partial t}&=-2\gamma(t)T_e(t)-2\frac{m_e}{m_i}{\gamma_{ei}}(t)T_e(t) \label{eq:TempEvolModeld}
\end{align}
where  $U_{ii}(t)$ is the average ion-ion correlation energy per particle in the plasma, which can be related to the spatial pair correlation function of the ions. \cite{kpp07} $U_{ii}$ is negative in UNPs.

The electron-ion equilibration terms in Eqs.\ \ref{eq:TempEvolModelc} and \ref{eq:TempEvolModeld}, which contain ${\gamma_{ei}}$ will be discussed below. When they are omitted and the correlation energy is neglected ($U_{ii}\equiv 0$), Eqs.\ \ref{eq:TempEvolModela}-\ref{eq:TempEvolModeld} can be solved exactly by the analytic solution presented in Eqs.\ \ref{eq:sizeevolution} and  \ref{eq:tempevolution}.

The inclusion of the correlation energy in Eqs.\ \ref{eq:TempEvolModela} - \ref{eq:TempEvolModeld} allows an approximate treatment of correlation effects such as disorder-induced heating. $U_{ii}$ can only be calculated with a full molecular dynamics description of the plasma. \cite{kpp07} It's evolution can be approximated, however, as \cite{ppr04}

\begin{equation}\label{eq:CorrelationEnergyEvolution}
  \frac{\partial U_{ii}(t)}{\partial t} = -\frac{U_{ii}(t)-U_{ii,eq}(t)}{\tau_{corr}(t)},
\end{equation}
where the timescale for relaxation of the correlation energy to its equilibrium value, $U_{ii,eq}$,  \cite{cp98} is taken as the inverse-ion plasma frequency $\tau_{corr}=\sqrt{m_i \varepsilon_0 /e^2 {n_i}}$.

At plasma formation, $U_{ii}$ is set to zero, but its magnitude increases rapidly, raising the ion temperature close to the equilibrium temperature $T_{DIH}$ on a timescale of $\tau_{corr}$. The  ion temperature after establishment of local equilibrium is not put into the model as an \textit{ad hoc} parameter; it emerges naturally from the dynamics. This treatment also includes the retardation of the expansion due to decreasing magnitude of $U_{ii}$ with decreasing density, which is a small effect for our conditions. The ion KEOs observed  experimentally  during the disorder-induced-heating phase are not described at this level of treatment. The slow approach to $T_{DIH}$ during the damping of the KEOs mentioned in Sec.\ \ref{Sec:DIHandKEO} is also not described.

Inherent in the traditional interpretation of Eq.\ \ref{eq:TempEvolModelc}, however, is an assumption of global thermal equilibrium for the ions that is not well-justified. The equilibrium correlation energy depends on the local density, and disorder-induced heating only leads to local thermal equilibrium. The characteristic timescale for global ion thermal equilibration $\tau_{therm}$ is set by the thermal conductivity for the strongly coupled ions, which can be approximated as $\lambda\approx n k_B \omega_{pi} a_{ws}^2$. \cite{dha04} Temperature equilibration between the ions in two regions of volume $V$, interface area $A$, and separation $d$ is $\sim c_V V d/\lambda A$, where $c_V=3nk_B/2$ is the ion heat capacity per volume. Taking all length scales as the plasma size $\sigma$, this yields $\tau_{therm}\approx\sigma^2/\omega_{pi}a_{ws}^2$, which is on the order of $10^{-3}$\,s for a typical UNP - much longer than other dynamic timescales for the plasma.

This creates a significant complication for a rigorous description of the plasma, but we take advantage of the extremely long timescale for ion global equilibration to make a significant approximation to describe our data. We restrict our analysis to a central region of the plasma in which the average density is 96\% of the peak density. We consider Eq.\ \ref{eq:TempEvolModelc} as describing an effective temperature for local thermal equilibribrium within this region. While previous treatments \cite{ppr04} have approximated $U_{ii,eq}(t)$ as an average over the equilibrium correlation energy in the entire plasma, we take it as the correlation energy for the measured plasma density, which is the average density of our region of interest. Several additional factors make Eq.\ \ref{eq:TempEvolModelc} a reasonable approximation for the local ion conditions. The actual error introduced with this approximation is small because the temperature after local equilibration $\sim T_{DIH}$ varies slowly with density.  Also, the self-similar plasma expansion (Eqs.\ \ref{eq:sizeevolution} and \ref{eq:tempevolution}) leads to the same relative change in volume and adiabatic cooling in all regions of the plasma. Finally, the plasma expansion, which controls the adiabatic cooling dynamics,  is dominated by the electron temperature for our conditions and is relatively insensitive to the ion temperature.



With this interpretation, the equilibrium value of the correlation energy ($U_{ii,eq}<0$) is estimated as \cite{cp98}
\begin{equation}\label{eq:EquilAverageCorrelationEnergy}
\begin{split}
 U_{ii,eq}(T_i,n)&=k_B T_i \Gamma^{3/2}(T_i,n)\\
 &\left(\frac{A_1}{\sqrt{A_2+\Gamma(T_i,n)}}+\frac{A_3}{1+\Gamma(T_i,n)}\right).
\end{split}
\end{equation}
where $T_i$ and $n$ are the instantaneous  local ion temperature and density, and $A_1=-0.9052$, $A_2=0.6322$, and $A_3=-\sqrt{3}/2-A_1/\sqrt{A_2}$.

Eqs.\ \ref{eq:TempEvolModela}-\ref{eq:CorrelationEnergyEvolution}
neglecting electron-ion equilibration qualitatively predict the ion temperature evolution, as shown by rescaling data by $T_{DIH}$ and $\tau_{exp}$ and overlaying the results of the model (solid lines in Figs.\ \ref{fig:Expansion} (b) and \ref{fig:ExpansionSizes} (b)). But quantitatively, the theory clearly diverges from the data after the DIH phase  and underestimates the ion temperature during the expansion. This implies that the model misses a significant amount of energy that is transferred to the ions during the evolution, and we now turn to a discussion of the sources of this energy.

\subsubsection{Electron-Ion Thermalization}
Collisional energy transfer between electrons and ions is typically neglected in theory and experiments on UNPs. \cite{kpp07} Due to the large  mass difference, complete thermalization between the two would require $10$ to $100$ times the expansion time ($\tau_{exp}$) of an UNP. Thus, the energy lost by  electrons due to collisions with ions is insignificant compared to the electron temperature, \cite{kkb99} and   the expansion of the plasma and the electron temperature evolution can be described by Eqs.\ \ref{eq:sizeevolution} and  \ref{eq:tempevolution}.

However, if one is concerned with the ion temperature, electron-ion thermalization must be considered since the ions are typically orders of magnitude colder than electrons. This is particularly important for proposed efforts to increase Coulomb coupling  by laser cooling \cite{kon02,ppr05JPB} or circumventing DIH \cite{glh10} because a small transfer of electron energy to thermal ion energy heats the ions significantly. In general, thermal relaxation in a plasma has been a long-standing fundamental interest \cite{daw64,mn70,bm86,hdm87,mar99,mn03,bfm06,dd08} as well as a critical aspect of many areas of study such as inertial confinement fusion, \cite{mdh08,ggm08,rfl09,vgb10,bsc12} warm dense matter, \cite{sbv10} and space plasmas. \cite{pap71,sm91,gyp01}

The measurements presented here are the first to follow the ion temperature on the timescale of the expansion and observe the effects of electron-ion thermalization in an UNP.  To describe the data, and help us identify the contribution of this heating mechanism, we start with the classic electron-ion collision frequency in a singly ionized plasmas, assuming $n_i = n_e$, $m_i >> m_e$ and $T_e >> T_i$, \cite{spi56}
\begin{eqnarray}
  \gamma_{ei} &=& \frac{4\sqrt{2\pi}n_i e^4 \ln[\Lambda]}{3 (4\pi\varepsilon_0)^2 m_e^{1/2}(k_B T_e)^{3/2}}  \nonumber \\
   &=& \sqrt{\frac{2}{3\pi}}\Gamma_e^{3/2}\omega_{pe}\ln[\Lambda] ,\label{eq:GammaEIC}
\end{eqnarray}
and the ion heating and electron cooling rates due to electron-ion collisions (EICs),
\begin{equation}
\label{eq:HeatingDiff2}
\frac{dT_i}{dt}=-\frac{dT_e}{dt}=2\frac{m_e}{m_i}{\gamma_{ei}}T_e.
\end{equation}
The argument of the Coulomb logarithm is $\Lambda={1/\sqrt{3}}{\Gamma_e^{3/2}}$. The Coulomb coupling parameter for the electrons is never greater than 1 for our experiments, so strong coupling effects are small, \cite{hmc83} as has been shown by comparison of Eq.\ \ref{eq:GammaEIC} with quantum T-matrix calculations \cite{gms02} in this regime of coupling.

%
The heating and cooling terms are included in the evolution of ion and electron temperatures (Eqs.\ \ref{eq:TempEvolModelc} and \ref{eq:TempEvolModeld}). We use the measured, average plasma density for the region of interest in these expressions. Following a similar line of reasoning as discussed in the context of the correlation energy, this provides a reasonable approximation of the evolution of the local ion temperature in the central region of the plasma because of the very slow global equilibration of the ions.

\begin{figure}[h]
\includegraphics{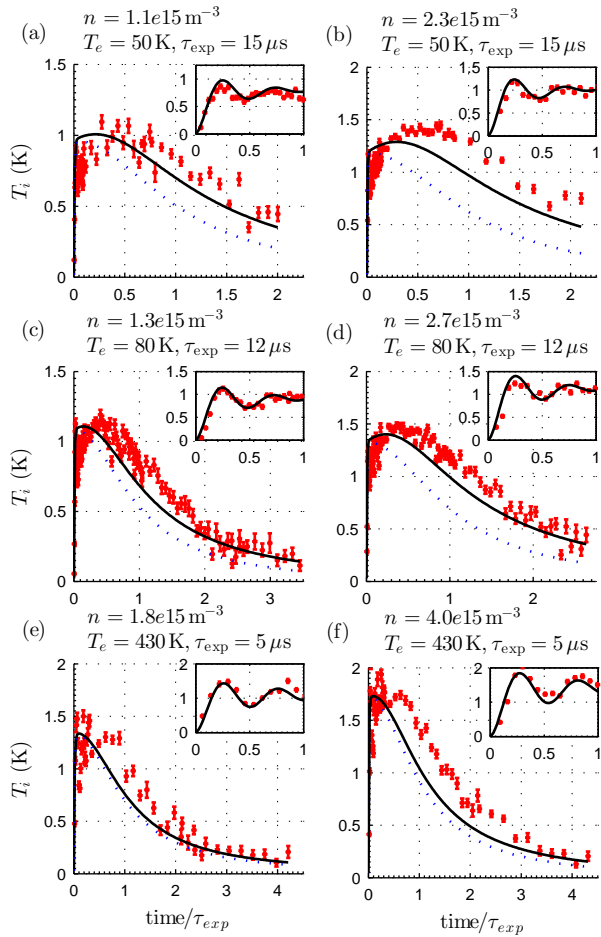}
\caption{\label{fig:EICs} Comparison of data with models including and not including electron ion collisions.
The lines are simulation results of the ion temperature model (Eqs.\ \ref{eq:TempEvolModela}-\ref{eq:CorrelationEnergyEvolution})
including correlation effects and cloud expansion \cite{ppr04}  with (solid) and without (dotted) electron-ion thermalization. The insets show early time KEOs with time normalized by $2\pi \omega_{pi}^{-1}$. The solid lines in the insets are results from MD simulations fit to the data in order to determine experimental density.}
\end{figure}

With the inclusion of electron-ion thermalization in the model, it is possible to determine the
significance of the effect for ion temperature evolution. Figure \ref{fig:EICs} shows data and simulation for various regimes of UNP parameters.
The simulation results show that the heating from EICs is more significant for higher density, smaller electron temperature, and larger initial plasma size. These conditions increase the collision rate and the time ($\sim\tau_{exp}$) before the density drops and collisions become negligible. In Fig.\ \ref{fig:EICs}(b), electron-ion thermalization contributes 0.5\,K and  doubles the ion temperature at later times.
For the opposite extreme of the accessible regime of these parameters in UNPs, the effect is very small.

%

\subsubsection{Additional Heating/Ion Acoustic Waves}

It is evident from Fig.\ \ref{fig:EICs} that the simulation still underestimates the measured ion temperature during the expansion by as much as several hundred millikelvin, and that the excess energy appears on the timescale of $\tau_{exp}$. We observe that this effect is very sensitive to experimental parameters such as laser alignment and beam quality. But in general  the additional energy is more significant for larger density and electron temperature. 

One possibility is that hydrodynamic expansion velocity is being misinterpreted as thermal velocity. The plasma is expanding with a velocity that varies in space, and LIF signal captured from a region with finite width $\delta x$ will reflect a spread of expansion velocities that can broaden the measured spectrum. For the self-similar expansion (Eqs.\ \ref{eq:sizeevolution}-\ref{eq:tempevolution}), the velocity spread $\delta v = \gamma(t) \delta x$ peaks at $t=\tau_{exp}$ and then decreases. Assuming an ideal single pixel region of width $\delta x = $ \SI{13}{\micro\metre}, $\delta v$ peaks at a value of \SI{1}{\metre\per\second} corresponding to a Doppler broadening of  \SI{2}{\mega\hertz} or  \SI{8}{\milli\kelvin} in temperature units. The actual value would increase if the expansion deviates from the ideal self-similar expansion or if the point spread function used  for image deconvolution is inaccurate or not  constant throughout the fluorescence volume. However, it is unlikely this accounts for a large fraction of the observed heating.

Another possibility is that Eqs. \ref{eq:GammaEIC} and \ref{eq:HeatingDiff2} underestimate the amount of ion heating due to electron-ion collisions. We do not attempt to use our data to extract the electron-ion thermalization rate because of the difficulty in separating this heat source from other possible mechanisms. While there are many assumptions that go into the derivation of Eqs. \ref{eq:GammaEIC} and \ref{eq:HeatingDiff2}, it is unlikely that this can be the source of the anomalous heating because it appears to scale very differently than heating due to electron-ion collisions (Fig. \ref{fig:EICs}).

We hypothesize that the largest contribution to the observed excess ion energy reflects ion acoustic waves (IAWs) that are excited during the ionization process due to deviations from a smooth Gaussian in the initial density distribution of the plasma.
Previously we have used transmission masks on the ionizing beam to intentionally modulate the initial density distribution leading to the excitation of IAWs. \cite{cmk10,mck11,kmo12} However, similar excitations would be caused by any imperfection in the ionizing laser profiles and/or atom density distribution. Long-wavelength IAW excitations with a broad spread of wavevectors would increase the RMS velocity width of the ions. Short wavelength excitations should decay quickly through Landau damping, increasing the ion thermal energy. Both would increase the temperature as measured by our LIF probe.

To assess the feasibility of this explanation, we estimate the energy per ion from a single IAW mode of frequency $\omega$, wavevector $k$, and  fractional density modulation $\frac{\delta n}{n_0}$
as \cite{kmo12}
\begin{equation}
\Delta E_{IAW} \approx \frac{1}{4} m_i\left(\frac{\omega}{k}\frac{\delta n}{n_0}\right)^2.
\end{equation}
 In the long wavelength limit ${\omega}/{k}$ can be replaced with the sound speed $\sqrt{k_B T_e/m_i}$. 
 If we assume all the IAW energy is transferred into thermal energy,
 we find that
\begin{equation}
\Delta T_{i,max} \approx \frac{1}{6} T_e \left(\frac{\delta n}{n_0}\right)^2.
\end{equation}
For a $\frac{\delta n}{n_0}$ of only 5\% and $T_e = $ \SI{430}{\kelvin} this is nearly \SI{200}{\milli\kelvin} of energy. 
In the short wavelength limit ${\omega}/{k}$ is replaced by ${\omega_{pi}}/{k}$ leading to a decreasing energy with wave vector,
\begin{equation}
\Delta T_{i,max} \approx \frac{1}{6} \frac{ne^2}{\epsilon_0 k_B k^2 }\left(\frac{\delta n}{n_0}\right)^2
\end{equation}

This effect seems capable of providing the observed heating, and residues of the density-distribution fit to a 2D Gaussian are often on the order of \SI{5}{\percent} on a $\sim$ \SI{1}{\milli\metre} length scale.
 Similar depth ion holes have been shown to perturb the ion temperatures by several hundred millikelvin. \cite{pck11,mcs13}

Despite great effort, deviation of the density distribution from an ideal Gaussian remains a limitation for achieving lower ion temperatures. For all but the lowest atom densities, the magneto-optical trap for the atoms is optically thick to the laser-cooling light, this results in non-Gaussian atom distributions that are inherited by the plasma. Also, the output of the pulsed-dye laser that ionizes the atoms has spatial intensity modulations that are imprinted onto the plasma. Complete saturation of the ionizing process could minimize this source of heat but it would require larger pulse energies than available with our current system.

\section{Increase in $\Gamma_{i}$ with Expansion}

 For the best chance of seeing increasing $\Gamma_i$ with expansion, \cite{ppr05PRL} minimal heating  and fast expansion is required. Fast expansion is achieved by small cloud size and high $T_e$. 
 Higher density doesn't change the initial coupling parameter, but it  increases the DIH temperature - minimizing the relative decrease in $\Gamma_i$ coming from  additional heating sources. Higher density also  enhances EIC  heating, but for sufficiently high electron temperatures, e.g. near the auto-ionizing resonance, the EIC cross section is so small that the highest achievable densities should not experience significant EIC heating. Higher density also allows data collection at later times. Therefore, at least for our accessible experimental parameters, high $T_e$, small cloud and high density are optimal for observing increasing $\Gamma_i$.

\begin{figure}[h]
\includegraphics{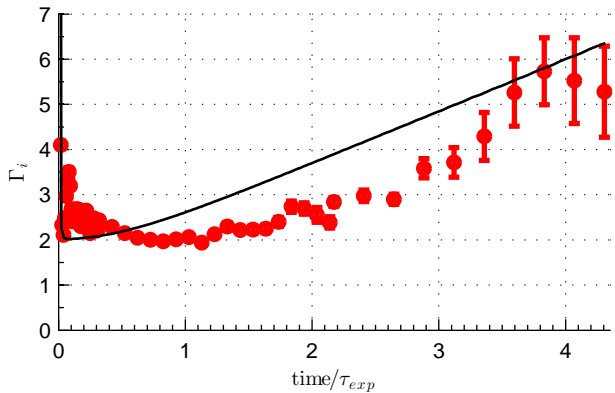}
\caption{\label{fig:Gamma} Increase in the ion Coulomb coupling parameter with plasma expansion for $T_e(0) =$ \SI{430}{\kelvin},  $n_0 = $ \SI{4e15}{\per\metre\cubed}, and $\sigma_0=$ \SI{1}{\milli\metre}. The line is the prediction of Eqs.\ \ref{eq:TempEvolModela}-\ref{eq:CorrelationEnergyEvolution}.}
\end{figure}

 Measurements at later times are limited by plummeting densities and signal-to-noise ratios, but for the best conditions, we see $\Gamma_i$ increase above 5, as shown in Fig. \ref{fig:Gamma}. EICs and other sources of heating prevent $\Gamma_i$ from reaching as high as expected from Eqs.\ \ref{eq:TempEvolModela}-\ref{eq:CorrelationEnergyEvolution} neglecting these effects. The measured value of $\Gamma_i$ is calculated from observed ion temperature and density, and we are not able to determine if spatial correlations exist corresponding to equilibrium at this value. \cite{ppr05PRL}

\section{Conclusion}
We have used spatially resolved fluorescence spectroscopy to measure ion temperature evolution during the expansion of an UNP for various initial plasma sizes, electron temperatures, and densities. We have observed several effects in UNPs for the first time, such as ion adiabatic cooling and ion heating due to collisions with electrons. We have presented a numerical model describing the evolution of plasma size and electron and ion temperatures including the effects of electron-ion collisions. We also presented evidence for additional ion heating, which we hypothesize is dominated by IAWs excited during plasma creation. We see increasing coupling for ions with time, but it is limited by these heating effects. Attempts to access stronger Coulomb coupling, such as through proposed schemes to laser cool ions in UNPs \cite{gks97,gks98,gks02,kon02,kag03,ppr04prl,ppr05JPB,kra09,gik09} or simply through plasma expansion as demonstrated here, should benefit from improved control of the initial density distribution of the plasma.

\begin{acknowledgments}
We thank Thomas Pohl for providing results from molecular dynamics simulation of equilibrating plasmas and Barry Dunning for the loan of a laser for LIF excitation. This work was supported by the Department of Energy and National Science Foundation (PHY-0714603) and the Air Force Office of Scientific Research (FA9550-12-1-0267).
\end{acknowledgments}

\bibliography{bibliography}

\begin{thebibliography}{94}%
\makeatletter
\providecommand \@ifxundefined [1]{%
 \@ifx{#1\undefined}
}%
\providecommand \@ifnum [1]{%
 \ifnum #1\expandafter \@firstoftwo
 \else \expandafter \@secondoftwo
 \fi
}%
\providecommand \@ifx [1]{%
 \ifx #1\expandafter \@firstoftwo
 \else \expandafter \@secondoftwo
 \fi
}%
\providecommand \natexlab [1]{#1}%
\providecommand \enquote  [1]{``#1''}%
\providecommand \bibnamefont  [1]{#1}%
\providecommand \bibfnamefont [1]{#1}%
\providecommand \citenamefont [1]{#1}%
\providecommand \href@noop [0]{\@secondoftwo}%
\providecommand \href [0]{\begingroup \@sanitize@url \@href}%
\providecommand \@href[1]{\@@startlink{#1}\@@href}%
\providecommand \@@href[1]{\endgroup#1\@@endlink}%
\providecommand \@sanitize@url [0]{\catcode `\\12\catcode `\$12\catcode
  `\&12\catcode `\#12\catcode `\^12\catcode `\_12\catcode `\%12\relax}%
\providecommand \@@startlink[1]{}%
\providecommand \@@endlink[0]{}%
\providecommand \url  [0]{\begingroup\@sanitize@url \@url }%
\providecommand \@url [1]{\endgroup\@href {#1}{\urlprefix }}%
\providecommand \urlprefix  [0]{URL }%
\providecommand \Eprint [0]{\href }%
\providecommand \doibase [0]{http://dx.doi.org/}%
\providecommand \selectlanguage [0]{\@gobble}%
\providecommand \bibinfo  [0]{\@secondoftwo}%
\providecommand \bibfield  [0]{\@secondoftwo}%
\providecommand \translation [1]{[#1]}%
\providecommand \BibitemOpen [0]{}%
\providecommand \bibitemStop [0]{}%
\providecommand \bibitemNoStop [0]{.\EOS\space}%
\providecommand \EOS [0]{\spacefactor3000\relax}%
\providecommand \BibitemShut  [1]{\csname bibitem#1\endcsname}%
\let\auto@bib@innerbib\@empty
\bibitem [{\citenamefont {Killian}\ \emph {et~al.}(2007)\citenamefont
  {Killian}, \citenamefont {Pattard}, \citenamefont {Pohl},\ and\ \citenamefont
  {Rost}}]{kpp07}%
  \BibitemOpen
  \bibfield  {author} {\bibinfo {author} {\bibfnamefont {T.~C.}\ \bibnamefont
  {Killian}}, \bibinfo {author} {\bibfnamefont {T.}~\bibnamefont {Pattard}},
  \bibinfo {author} {\bibfnamefont {T.}~\bibnamefont {Pohl}}, \ and\ \bibinfo
  {author} {\bibfnamefont {J.~M.}\ \bibnamefont {Rost}},\ }\href@noop {}
  {\bibfield  {journal} {\bibinfo  {journal} {{ Physics Reports}}\ }\textbf
  {\bibinfo {volume} {449}},\ \bibinfo {pages} {77} (\bibinfo {year}
  {2007})}\BibitemShut {NoStop}%
\bibitem [{\citenamefont {Murillo}(2001)}]{mur01}%
  \BibitemOpen
  \bibfield  {author} {\bibinfo {author} {\bibfnamefont {M.~S.}\ \bibnamefont
  {Murillo}},\ }\href@noop {} {\bibfield  {journal} {\bibinfo  {journal}
  {{Phys. Rev. Lett.}}\ }\textbf {\bibinfo {volume} {87}},\ \bibinfo {pages}
  {115003} (\bibinfo {year} {2001})}\BibitemShut {NoStop}%
\bibitem [{\citenamefont {Kuzmin}\ and\ \citenamefont
  {O'Neil}(2002{\natexlab{a}})}]{kon02prl}%
  \BibitemOpen
  \bibfield  {author} {\bibinfo {author} {\bibfnamefont {S.~G.}\ \bibnamefont
  {Kuzmin}}\ and\ \bibinfo {author} {\bibfnamefont {T.~M.}\ \bibnamefont
  {O'Neil}},\ }\href@noop {} {\bibfield  {journal} {\bibinfo  {journal} {{Phys.
  Rev. Lett.}}\ }\textbf {\bibinfo {volume} {88}},\ \bibinfo {pages} {65003}
  (\bibinfo {year} {2002}{\natexlab{a}})}\BibitemShut {NoStop}%
\bibitem [{\citenamefont {Mazevet}, \citenamefont {Collins},\ and\
  \citenamefont {Kress}(2002)}]{mck02}%
  \BibitemOpen
  \bibfield  {author} {\bibinfo {author} {\bibfnamefont {S.}~\bibnamefont
  {Mazevet}}, \bibinfo {author} {\bibfnamefont {L.~A.}\ \bibnamefont
  {Collins}}, \ and\ \bibinfo {author} {\bibfnamefont {J.~D.}\ \bibnamefont
  {Kress}},\ }\href@noop {} {\bibfield  {journal} {\bibinfo  {journal} {{Phys.
  Rev. Lett.}}\ }\textbf {\bibinfo {volume} {88}},\ \bibinfo {pages} {55001}
  (\bibinfo {year} {2002})}\BibitemShut {NoStop}%
\bibitem [{\citenamefont {Chen}\ \emph {et~al.}(2004)\citenamefont {Chen},
  \citenamefont {Simien}, \citenamefont {Laha}, \citenamefont {Gupta},
  \citenamefont {Martinez}, \citenamefont {Mickelson}, \citenamefont {Nagel},\
  and\ \citenamefont {Killian}}]{csl04}%
  \BibitemOpen
  \bibfield  {author} {\bibinfo {author} {\bibfnamefont {Y.~C.}\ \bibnamefont
  {Chen}}, \bibinfo {author} {\bibfnamefont {C.~E.}\ \bibnamefont {Simien}},
  \bibinfo {author} {\bibfnamefont {S.}~\bibnamefont {Laha}}, \bibinfo {author}
  {\bibfnamefont {P.}~\bibnamefont {Gupta}}, \bibinfo {author} {\bibfnamefont
  {Y.~N.}\ \bibnamefont {Martinez}}, \bibinfo {author} {\bibfnamefont {P.~G.}\
  \bibnamefont {Mickelson}}, \bibinfo {author} {\bibfnamefont {S.~B.}\
  \bibnamefont {Nagel}}, \ and\ \bibinfo {author} {\bibfnamefont {T.~C.}\
  \bibnamefont {Killian}},\ }\href@noop {} {\bibfield  {journal} {\bibinfo
  {journal} {{Phys. Rev. Lett.}}\ }\textbf {\bibinfo {volume} {93}},\ \bibinfo
  {pages} {265003} (\bibinfo {year} {2004})}\BibitemShut {NoStop}%
\bibitem [{\citenamefont {Cummings}\ \emph {et~al.}(2005)\citenamefont
  {Cummings}, \citenamefont {Daily}, \citenamefont {Durfee},\ and\
  \citenamefont {Bergeson}}]{cdd05}%
  \BibitemOpen
  \bibfield  {author} {\bibinfo {author} {\bibfnamefont {E.~A.}\ \bibnamefont
  {Cummings}}, \bibinfo {author} {\bibfnamefont {J.~E.}\ \bibnamefont {Daily}},
  \bibinfo {author} {\bibfnamefont {D.~S.}\ \bibnamefont {Durfee}}, \ and\
  \bibinfo {author} {\bibfnamefont {S.~D.}\ \bibnamefont {Bergeson}},\
  }\href@noop {} {\bibfield  {journal} {\bibinfo  {journal} {{Phys. Rev.
  Lett.}}\ }\textbf {\bibinfo {volume} {95}},\ \bibinfo {pages} {235001}
  (\bibinfo {year} {2005})}\BibitemShut {NoStop}%
\bibitem [{\citenamefont {Bergeson}\ \emph {et~al.}(2011)\citenamefont
  {Bergeson}, \citenamefont {Denning}, \citenamefont {Lyon},\ and\
  \citenamefont {Robicheaux}}]{bdl11}%
  \BibitemOpen
  \bibfield  {author} {\bibinfo {author} {\bibfnamefont {S.~D.}\ \bibnamefont
  {Bergeson}}, \bibinfo {author} {\bibfnamefont {A.}~\bibnamefont {Denning}},
  \bibinfo {author} {\bibfnamefont {M.}~\bibnamefont {Lyon}}, \ and\ \bibinfo
  {author} {\bibfnamefont {F.}~\bibnamefont {Robicheaux}},\ }\href {\doibase
  10.1103/PhysRevA.83.023409} {\bibfield  {journal} {\bibinfo  {journal} {Phys.
  Rev. A}\ }\textbf {\bibinfo {volume} {83}},\ \bibinfo {pages} {023409}
  (\bibinfo {year} {2011})}\BibitemShut {NoStop}%
\bibitem [{\citenamefont {Lyon}, \citenamefont {Bergeson},\ and\ \citenamefont
  {Murillo}(2013)}]{lbm13}%
  \BibitemOpen
  \bibfield  {author} {\bibinfo {author} {\bibfnamefont {M.}~\bibnamefont
  {Lyon}}, \bibinfo {author} {\bibfnamefont {S.~D.}\ \bibnamefont {Bergeson}},
  \ and\ \bibinfo {author} {\bibfnamefont {M.~S.}\ \bibnamefont {Murillo}},\
  }\href {\doibase 10.1103/PhysRevE.87.033101} {\bibfield  {journal} {\bibinfo
  {journal} {Phys. Rev. E}\ }\textbf {\bibinfo {volume} {87}},\ \bibinfo
  {pages} {033101} (\bibinfo {year} {2013})}\BibitemShut {NoStop}%
\bibitem [{\citenamefont {Bannasch}\ \emph {et~al.}(2012)\citenamefont
  {Bannasch}, \citenamefont {Castro}, \citenamefont {McQuillen}, \citenamefont
  {Pohl},\ and\ \citenamefont {Killian}}]{bcm12}%
  \BibitemOpen
  \bibfield  {author} {\bibinfo {author} {\bibfnamefont {G.}~\bibnamefont
  {Bannasch}}, \bibinfo {author} {\bibfnamefont {J.}~\bibnamefont {Castro}},
  \bibinfo {author} {\bibfnamefont {P.}~\bibnamefont {McQuillen}}, \bibinfo
  {author} {\bibfnamefont {T.}~\bibnamefont {Pohl}}, \ and\ \bibinfo {author}
  {\bibfnamefont {T.~C.}\ \bibnamefont {Killian}},\ }\href {\doibase
  10.1103/PhysRevLett.109.185008} {\bibfield  {journal} {\bibinfo  {journal}
  {Phys. Rev. Lett.}\ }\textbf {\bibinfo {volume} {109}},\ \bibinfo {pages}
  {185008} (\bibinfo {year} {2012})}\BibitemShut {NoStop}%
\bibitem [{\citenamefont {Ichimaru}(1982)}]{ich82}%
  \BibitemOpen
  \bibfield  {author} {\bibinfo {author} {\bibfnamefont {S.}~\bibnamefont
  {Ichimaru}},\ }\href@noop {} {\bibfield  {journal} {\bibinfo  {journal}
  {{Reviews of Modern Physics}}\ }\textbf {\bibinfo {volume} {54}},\ \bibinfo
  {pages} {1017} (\bibinfo {year} {1982})}\BibitemShut {NoStop}%
\bibitem [{\citenamefont {Murillo}(2007)}]{mur07pop}%
  \BibitemOpen
  \bibfield  {author} {\bibinfo {author} {\bibfnamefont {M.~S.}\ \bibnamefont
  {Murillo}},\ }\href {\doibase 10.1063/1.2436853} {\bibfield  {journal}
  {\bibinfo  {journal} {Phys. Plasmas}\ }\textbf {\bibinfo {volume} {14}},\
  \bibinfo {pages} {055702} (\bibinfo {year} {2007})}\BibitemShut {NoStop}%
\bibitem [{\citenamefont {Maxson}\ \emph {et~al.}(2013)\citenamefont {Maxson},
  \citenamefont {Bazarov}, \citenamefont {Wan}, \citenamefont {Padmore},\ and\
  \citenamefont {Coleman-Smith}}]{mbw13}%
  \BibitemOpen
  \bibfield  {author} {\bibinfo {author} {\bibfnamefont {J.~M.}\ \bibnamefont
  {Maxson}}, \bibinfo {author} {\bibfnamefont {I.~V.}\ \bibnamefont {Bazarov}},
  \bibinfo {author} {\bibfnamefont {W.}~\bibnamefont {Wan}}, \bibinfo {author}
  {\bibfnamefont {H.~A.}\ \bibnamefont {Padmore}}, \ and\ \bibinfo {author}
  {\bibfnamefont {C.~E.}\ \bibnamefont {Coleman-Smith}},\ }\href
  {http://stacks.iop.org/1367-2630/15/i=10/a=103024} {\bibfield  {journal}
  {\bibinfo  {journal} {New Journal of Physics}\ }\textbf {\bibinfo {volume}
  {15}},\ \bibinfo {pages} {103024} (\bibinfo {year} {2013})}\BibitemShut
  {NoStop}%
\bibitem [{\citenamefont {Kulin}\ \emph {et~al.}(2000)\citenamefont {Kulin},
  \citenamefont {Killian}, \citenamefont {Bergeson},\ and\ \citenamefont
  {Rolston}}]{kkb00}%
  \BibitemOpen
  \bibfield  {author} {\bibinfo {author} {\bibfnamefont {S.}~\bibnamefont
  {Kulin}}, \bibinfo {author} {\bibfnamefont {T.~C.}\ \bibnamefont {Killian}},
  \bibinfo {author} {\bibfnamefont {S.~D.}\ \bibnamefont {Bergeson}}, \ and\
  \bibinfo {author} {\bibfnamefont {S.~L.}\ \bibnamefont {Rolston}},\
  }\href@noop {} {\bibfield  {journal} {\bibinfo  {journal} {{Phys. Rev.
  Lett.}}\ }\textbf {\bibinfo {volume} {85}},\ \bibinfo {pages} {318} (\bibinfo
  {year} {2000})}\BibitemShut {NoStop}%
\bibitem [{\citenamefont {Robicheaux}\ and\ \citenamefont
  {Hanson}(2002)}]{rha02}%
  \BibitemOpen
  \bibfield  {author} {\bibinfo {author} {\bibfnamefont {F.}~\bibnamefont
  {Robicheaux}}\ and\ \bibinfo {author} {\bibfnamefont {J.~D.}\ \bibnamefont
  {Hanson}},\ }\href@noop {} {\bibfield  {journal} {\bibinfo  {journal} {{Phys.
  Rev. Lett.}}\ }\textbf {\bibinfo {volume} {88}},\ \bibinfo {pages} {55002}
  (\bibinfo {year} {2002})}\BibitemShut {NoStop}%
\bibitem [{\citenamefont {Laha}\ \emph {et~al.}(2007)\citenamefont {Laha},
  \citenamefont {Gupta}, \citenamefont {Simien}, \citenamefont {Gao},
  \citenamefont {Castro},\ and\ \citenamefont {Killian}}]{lgs07}%
  \BibitemOpen
  \bibfield  {author} {\bibinfo {author} {\bibfnamefont {S.}~\bibnamefont
  {Laha}}, \bibinfo {author} {\bibfnamefont {P.}~\bibnamefont {Gupta}},
  \bibinfo {author} {\bibfnamefont {C.~E.}\ \bibnamefont {Simien}}, \bibinfo
  {author} {\bibfnamefont {H.}~\bibnamefont {Gao}}, \bibinfo {author}
  {\bibfnamefont {J.}~\bibnamefont {Castro}}, \ and\ \bibinfo {author}
  {\bibfnamefont {T.~C.}\ \bibnamefont {Killian}},\ }\href@noop {} {\bibfield
  {journal} {\bibinfo  {journal} {{Phys. Rev. Lett.}}\ }\textbf {\bibinfo
  {volume} {99}},\ \bibinfo {pages} {155001} (\bibinfo {year}
  {2007})}\BibitemShut {NoStop}%
\bibitem [{\citenamefont {Castro}, \citenamefont {McQuillen},\ and\
  \citenamefont {Killian}(2010)}]{cmk10}%
  \BibitemOpen
  \bibfield  {author} {\bibinfo {author} {\bibfnamefont {J.}~\bibnamefont
  {Castro}}, \bibinfo {author} {\bibfnamefont {P.}~\bibnamefont {McQuillen}}, \
  and\ \bibinfo {author} {\bibfnamefont {T.~C.}\ \bibnamefont {Killian}},\
  }\href@noop {} {\bibfield  {journal} {\bibinfo  {journal} {{Phys. Rev.
  Lett.}}\ }\textbf {\bibinfo {volume} {105}},\ \bibinfo {pages} {065004}
  (\bibinfo {year} {2010})}\BibitemShut {NoStop}%
\bibitem [{\citenamefont {McQuillen}\ \emph {et~al.}(2013)\citenamefont
  {McQuillen}, \citenamefont {Castro}, \citenamefont {Strickler}, \citenamefont
  {Bradshaw},\ and\ \citenamefont {Killian}}]{mcs13}%
  \BibitemOpen
  \bibfield  {author} {\bibinfo {author} {\bibfnamefont {P.}~\bibnamefont
  {McQuillen}}, \bibinfo {author} {\bibfnamefont {J.}~\bibnamefont {Castro}},
  \bibinfo {author} {\bibfnamefont {T.}~\bibnamefont {Strickler}}, \bibinfo
  {author} {\bibfnamefont {S.~J.}\ \bibnamefont {Bradshaw}}, \ and\ \bibinfo
  {author} {\bibfnamefont {T.~C.}\ \bibnamefont {Killian}},\ }\href {\doibase
  http://dx.doi.org/10.1063/1.4802813} {\bibfield  {journal} {\bibinfo
  {journal} {Phys. Plasmas}\ }\textbf {\bibinfo {volume} {20}},\ \bibinfo {eid}
  {043516} (\bibinfo {year} {2013})}\BibitemShut {NoStop}%
\bibitem [{\citenamefont {Twedt}\ and\ \citenamefont {Rolston}(2012)}]{tro12}%
  \BibitemOpen
  \bibfield  {author} {\bibinfo {author} {\bibfnamefont {K.~A.}\ \bibnamefont
  {Twedt}}\ and\ \bibinfo {author} {\bibfnamefont {S.~L.}\ \bibnamefont
  {Rolston}},\ }\href {\doibase 10.1103/PhysRevLett.108.065003} {\bibfield
  {journal} {\bibinfo  {journal} {Phys. Rev. Lett.}\ }\textbf {\bibinfo
  {volume} {108}},\ \bibinfo {pages} {065003} (\bibinfo {year}
  {2012})}\BibitemShut {NoStop}%
\bibitem [{\citenamefont {Lyubonko}, \citenamefont {Pohl},\ and\ \citenamefont
  {Rost}(2012)}]{lpr12}%
  \BibitemOpen
  \bibfield  {author} {\bibinfo {author} {\bibfnamefont {A.}~\bibnamefont
  {Lyubonko}}, \bibinfo {author} {\bibfnamefont {T.}~\bibnamefont {Pohl}}, \
  and\ \bibinfo {author} {\bibfnamefont {J.-M.}\ \bibnamefont {Rost}},\ }\href
  {http://stacks.iop.org/1367-2630/14/i=5/a=053039} {\bibfield  {journal}
  {\bibinfo  {journal} {New Journal of Physics}\ }\textbf {\bibinfo {volume}
  {14}},\ \bibinfo {pages} {053039} (\bibinfo {year} {2012})}\BibitemShut
  {NoStop}%
\bibitem [{\citenamefont {Killian}\ \emph {et~al.}(2001)\citenamefont
  {Killian}, \citenamefont {Lim}, \citenamefont {Kulin}, \citenamefont {Dumke},
  \citenamefont {Bergeson},\ and\ \citenamefont {Rolston}}]{klk01}%
  \BibitemOpen
  \bibfield  {author} {\bibinfo {author} {\bibfnamefont {T.~C.}\ \bibnamefont
  {Killian}}, \bibinfo {author} {\bibfnamefont {M.~J.}\ \bibnamefont {Lim}},
  \bibinfo {author} {\bibfnamefont {S.}~\bibnamefont {Kulin}}, \bibinfo
  {author} {\bibfnamefont {R.}~\bibnamefont {Dumke}}, \bibinfo {author}
  {\bibfnamefont {S.~D.}\ \bibnamefont {Bergeson}}, \ and\ \bibinfo {author}
  {\bibfnamefont {S.~L.}\ \bibnamefont {Rolston}},\ }\href@noop {} {\bibfield
  {journal} {\bibinfo  {journal} {{Phys. Rev. Lett.}}\ }\textbf {\bibinfo
  {volume} {86}},\ \bibinfo {pages} {3759} (\bibinfo {year}
  {2001})}\BibitemShut {NoStop}%
\bibitem [{\citenamefont {Bergeson}\ and\ \citenamefont
  {Robicheaux}(2008)}]{bro08}%
  \BibitemOpen
  \bibfield  {author} {\bibinfo {author} {\bibfnamefont {S.~D.}\ \bibnamefont
  {Bergeson}}\ and\ \bibinfo {author} {\bibfnamefont {F.}~\bibnamefont
  {Robicheaux}},\ }\href@noop {} {\bibfield  {journal} {\bibinfo  {journal}
  {Phys. Rev. Lett.}\ }\textbf {\bibinfo {volume} {101}},\ \bibinfo {pages}
  {073202} (\bibinfo {year} {2008})}\BibitemShut {NoStop}%
\bibitem [{\citenamefont {Robinson}\ \emph {et~al.}(2000)\citenamefont
  {Robinson}, \citenamefont {Tolra}, \citenamefont {Noel}, \citenamefont
  {Gallagher},\ and\ \citenamefont {Pillet}}]{rtn00}%
  \BibitemOpen
  \bibfield  {author} {\bibinfo {author} {\bibfnamefont {M.~P.}\ \bibnamefont
  {Robinson}}, \bibinfo {author} {\bibfnamefont {B.~L.}\ \bibnamefont {Tolra}},
  \bibinfo {author} {\bibfnamefont {M.~W.}\ \bibnamefont {Noel}}, \bibinfo
  {author} {\bibfnamefont {T.~F.}\ \bibnamefont {Gallagher}}, \ and\ \bibinfo
  {author} {\bibfnamefont {P.}~\bibnamefont {Pillet}},\ }\href@noop {}
  {\bibfield  {journal} {\bibinfo  {journal} {{Phys. Rev. Lett.}}\ }\textbf
  {\bibinfo {volume} {85}},\ \bibinfo {pages} {4466} (\bibinfo {year}
  {2000})}\BibitemShut {NoStop}%
\bibitem [{\citenamefont {Perry}\ and\ \citenamefont {Mourou}(1994)}]{pmo94}%
  \BibitemOpen
  \bibfield  {author} {\bibinfo {author} {\bibfnamefont {M.~D.}\ \bibnamefont
  {Perry}}\ and\ \bibinfo {author} {\bibfnamefont {G.}~\bibnamefont {Mourou}},\
  }\href@noop {} {\bibfield  {journal} {\bibinfo  {journal} {{Science}}\
  }\textbf {\bibinfo {volume} {264}},\ \bibinfo {pages} {917} (\bibinfo {year}
  {1994})}\BibitemShut {NoStop}%
\bibitem [{\citenamefont {Lindl}(1995)}]{lin95}%
  \BibitemOpen
  \bibfield  {author} {\bibinfo {author} {\bibfnamefont {J.}~\bibnamefont
  {Lindl}},\ }\href@noop {} {\bibfield  {journal} {\bibinfo  {journal} {{ Phys.
  Plasmas}}\ }\textbf {\bibinfo {volume} {2}},\ \bibinfo {pages} {3933}
  (\bibinfo {year} {1995})}\BibitemShut {NoStop}%
\bibitem [{\citenamefont {Daido}(2002)}]{dai02}%
  \BibitemOpen
  \bibfield  {author} {\bibinfo {author} {\bibfnamefont {H.}~\bibnamefont
  {Daido}},\ }\href@noop {} {\bibfield  {journal} {\bibinfo  {journal}
  {{Reports on Progress in Physics}}\ }\textbf {\bibinfo {volume} {65}},\
  \bibinfo {pages} {1513} (\bibinfo {year} {2002})}\BibitemShut {NoStop}%
\bibitem [{\citenamefont {Clark}\ \emph {et~al.}(2000)\citenamefont {Clark},
  \citenamefont {Krushelnick}, \citenamefont {Zepf}, \citenamefont {Beg},
  \citenamefont {Tatarakis}, \citenamefont {Machacek}, \citenamefont {Santala},
  \citenamefont {Watts}, \citenamefont {Norreys},\ and\ \citenamefont
  {Dangor}}]{ckz00}%
  \BibitemOpen
  \bibfield  {author} {\bibinfo {author} {\bibfnamefont {E.~L.}\ \bibnamefont
  {Clark}}, \bibinfo {author} {\bibfnamefont {K.}~\bibnamefont {Krushelnick}},
  \bibinfo {author} {\bibfnamefont {M.}~\bibnamefont {Zepf}}, \bibinfo {author}
  {\bibfnamefont {F.~N.}\ \bibnamefont {Beg}}, \bibinfo {author} {\bibfnamefont
  {M.}~\bibnamefont {Tatarakis}}, \bibinfo {author} {\bibfnamefont
  {A.}~\bibnamefont {Machacek}}, \bibinfo {author} {\bibfnamefont {M.~I.~K.}\
  \bibnamefont {Santala}}, \bibinfo {author} {\bibfnamefont {I.}~\bibnamefont
  {Watts}}, \bibinfo {author} {\bibfnamefont {P.~A.}\ \bibnamefont {Norreys}},
  \ and\ \bibinfo {author} {\bibfnamefont {A.~E.}\ \bibnamefont {Dangor}},\
  }\href {\doibase 10.1103/PhysRevLett.85.1654} {\bibfield  {journal} {\bibinfo
   {journal} {Phys. Rev. Lett.}\ }\textbf {\bibinfo {volume} {85}},\ \bibinfo
  {pages} {1654} (\bibinfo {year} {2000})}\BibitemShut {NoStop}%
\bibitem [{\citenamefont {Snavely}\ \emph {et~al.}(2000)\citenamefont
  {Snavely}, \citenamefont {Key}, \citenamefont {Hatchett}, \citenamefont
  {Cowan}, \citenamefont {Roth}, \citenamefont {Phillips}, \citenamefont
  {Stoyer}, \citenamefont {Henry}, \citenamefont {Sangster}, \citenamefont
  {Singh}, \citenamefont {Wilks}, \citenamefont {MacKinnon}, \citenamefont
  {Offenberger}, \citenamefont {Pennington}, \citenamefont {Yasuike},
  \citenamefont {Langdon}, \citenamefont {Lasinski}, \citenamefont {Johnson},
  \citenamefont {Perry},\ and\ \citenamefont {Campbell}}]{skh00}%
  \BibitemOpen
  \bibfield  {author} {\bibinfo {author} {\bibfnamefont {R.~A.}\ \bibnamefont
  {Snavely}}, \bibinfo {author} {\bibfnamefont {M.~H.}\ \bibnamefont {Key}},
  \bibinfo {author} {\bibfnamefont {S.~P.}\ \bibnamefont {Hatchett}}, \bibinfo
  {author} {\bibfnamefont {T.~E.}\ \bibnamefont {Cowan}}, \bibinfo {author}
  {\bibfnamefont {M.}~\bibnamefont {Roth}}, \bibinfo {author} {\bibfnamefont
  {T.~W.}\ \bibnamefont {Phillips}}, \bibinfo {author} {\bibfnamefont {M.~A.}\
  \bibnamefont {Stoyer}}, \bibinfo {author} {\bibfnamefont {E.~A.}\
  \bibnamefont {Henry}}, \bibinfo {author} {\bibfnamefont {T.~C.}\ \bibnamefont
  {Sangster}}, \bibinfo {author} {\bibfnamefont {M.~S.}\ \bibnamefont {Singh}},
  \bibinfo {author} {\bibfnamefont {S.~C.}\ \bibnamefont {Wilks}}, \bibinfo
  {author} {\bibfnamefont {A.}~\bibnamefont {MacKinnon}}, \bibinfo {author}
  {\bibfnamefont {A.}~\bibnamefont {Offenberger}}, \bibinfo {author}
  {\bibfnamefont {D.~M.}\ \bibnamefont {Pennington}}, \bibinfo {author}
  {\bibfnamefont {K.}~\bibnamefont {Yasuike}}, \bibinfo {author} {\bibfnamefont
  {A.~B.}\ \bibnamefont {Langdon}}, \bibinfo {author} {\bibfnamefont {B.~F.}\
  \bibnamefont {Lasinski}}, \bibinfo {author} {\bibfnamefont {J.}~\bibnamefont
  {Johnson}}, \bibinfo {author} {\bibfnamefont {M.~D.}\ \bibnamefont {Perry}},
  \ and\ \bibinfo {author} {\bibfnamefont {E.~M.}\ \bibnamefont {Campbell}},\
  }\href {\doibase 10.1103/PhysRevLett.85.2945} {\bibfield  {journal} {\bibinfo
   {journal} {Phys. Rev. Lett.}\ }\textbf {\bibinfo {volume} {85}},\ \bibinfo
  {pages} {2945} (\bibinfo {year} {2000})}\BibitemShut {NoStop}%
\bibitem [{\citenamefont {Maksimchuk}\ \emph {et~al.}(2000)\citenamefont
  {Maksimchuk}, \citenamefont {Gu}, \citenamefont {Flippo}, \citenamefont
  {Umstadter},\ and\ \citenamefont {Bychenkov}}]{mgf00}%
  \BibitemOpen
  \bibfield  {author} {\bibinfo {author} {\bibfnamefont {A.}~\bibnamefont
  {Maksimchuk}}, \bibinfo {author} {\bibfnamefont {S.}~\bibnamefont {Gu}},
  \bibinfo {author} {\bibfnamefont {K.}~\bibnamefont {Flippo}}, \bibinfo
  {author} {\bibfnamefont {D.}~\bibnamefont {Umstadter}}, \ and\ \bibinfo
  {author} {\bibfnamefont {V.~Y.}\ \bibnamefont {Bychenkov}},\ }\href@noop {}
  {\bibfield  {journal} {\bibinfo  {journal} {Phys. Rev. Lett.}\ }\textbf
  {\bibinfo {volume} {84}},\ \bibinfo {pages} {4108} (\bibinfo {year}
  {2000})}\BibitemShut {NoStop}%
\bibitem [{\citenamefont {Symes}\ \emph {et~al.}(2007)\citenamefont {Symes},
  \citenamefont {Hohenberger}, \citenamefont {Henig},\ and\ \citenamefont
  {Ditmire}}]{shh07}%
  \BibitemOpen
  \bibfield  {author} {\bibinfo {author} {\bibfnamefont {D.~R.}\ \bibnamefont
  {Symes}}, \bibinfo {author} {\bibfnamefont {M.}~\bibnamefont {Hohenberger}},
  \bibinfo {author} {\bibfnamefont {A.}~\bibnamefont {Henig}}, \ and\ \bibinfo
  {author} {\bibfnamefont {T.}~\bibnamefont {Ditmire}},\ }\href@noop {}
  {\bibfield  {journal} {\bibinfo  {journal} {Phys. Rev. Lett.}\ }\textbf
  {\bibinfo {volume} {98}},\ \bibinfo {eid} {123401} (\bibinfo {year}
  {2007})}\BibitemShut {NoStop}%
\bibitem [{\citenamefont {Robicheaux}\ and\ \citenamefont
  {Hanson}(2003)}]{rha03}%
  \BibitemOpen
  \bibfield  {author} {\bibinfo {author} {\bibfnamefont {F.}~\bibnamefont
  {Robicheaux}}\ and\ \bibinfo {author} {\bibfnamefont {J.~D.}\ \bibnamefont
  {Hanson}},\ }\href@noop {} {\bibfield  {journal} {\bibinfo  {journal} {{
  Phys. Plasmas}}\ }\textbf {\bibinfo {volume} {10}},\ \bibinfo {pages} {2217}
  (\bibinfo {year} {2003})}\BibitemShut {NoStop}%
\bibitem [{\citenamefont {Baitin}\ and\ \citenamefont
  {Kuzanyan}(1998)}]{bku98}%
  \BibitemOpen
  \bibfield  {author} {\bibinfo {author} {\bibfnamefont {A.~V.}\ \bibnamefont
  {Baitin}}\ and\ \bibinfo {author} {\bibfnamefont {K.~M.}\ \bibnamefont
  {Kuzanyan}},\ }\href@noop {} {\bibfield  {journal} {\bibinfo  {journal} {J.
  Plasma Phys.}\ }\textbf {\bibinfo {volume} {59}},\ \bibinfo {pages} {83}
  (\bibinfo {year} {1998})}\BibitemShut {NoStop}%
\bibitem [{\citenamefont {Dorozhkina}\ and\ \citenamefont
  {Semenov}(1998)}]{dse98}%
  \BibitemOpen
  \bibfield  {author} {\bibinfo {author} {\bibfnamefont {D.~S.}\ \bibnamefont
  {Dorozhkina}}\ and\ \bibinfo {author} {\bibfnamefont {V.~E.}\ \bibnamefont
  {Semenov}},\ }\href@noop {} {\bibfield  {journal} {\bibinfo  {journal}
  {{Phys. Rev. Lett.}}\ }\textbf {\bibinfo {volume} {81}},\ \bibinfo {pages}
  {2691} (\bibinfo {year} {1998})}\BibitemShut {NoStop}%
\bibitem [{\citenamefont {Kovalev}, \citenamefont {Bychenkov},\ and\
  \citenamefont {Tikhonchuk}(2002)}]{kbt02}%
  \BibitemOpen
  \bibfield  {author} {\bibinfo {author} {\bibfnamefont {V.}~\bibnamefont
  {Kovalev}}, \bibinfo {author} {\bibfnamefont {V.}~\bibnamefont {Bychenkov}},
  \ and\ \bibinfo {author} {\bibfnamefont {V.}~\bibnamefont {Tikhonchuk}},\
  }\href {\doibase 10.1134/1.1506430} {\bibfield  {journal} {\bibinfo
  {journal} {Journal of Experimental and Theoretical Physics}\ }\textbf
  {\bibinfo {volume} {95}},\ \bibinfo {pages} {226} (\bibinfo {year}
  {2002})}\BibitemShut {NoStop}%
\bibitem [{\citenamefont {Kovalev}\ and\ \citenamefont
  {Bychenkov}(2003)}]{kby03}%
  \BibitemOpen
  \bibfield  {author} {\bibinfo {author} {\bibfnamefont {V.~F.}\ \bibnamefont
  {Kovalev}}\ and\ \bibinfo {author} {\bibfnamefont {V.~Y.}\ \bibnamefont
  {Bychenkov}},\ }\href@noop {} {\bibfield  {journal} {\bibinfo  {journal}
  {Phys. Rev. Lett.}\ }\textbf {\bibinfo {volume} {90}},\ \bibinfo {eid}
  {185004} (\bibinfo {year} {2003})}\BibitemShut {NoStop}%
\bibitem [{\citenamefont {Gupta}\ \emph {et~al.}(2007)\citenamefont {Gupta},
  \citenamefont {Laha}, \citenamefont {Simien}, \citenamefont {Gao},
  \citenamefont {Castro}, \citenamefont {Killian},\ and\ \citenamefont
  {Pohl}}]{gls07}%
  \BibitemOpen
  \bibfield  {author} {\bibinfo {author} {\bibfnamefont {P.}~\bibnamefont
  {Gupta}}, \bibinfo {author} {\bibfnamefont {S.}~\bibnamefont {Laha}},
  \bibinfo {author} {\bibfnamefont {C.~E.}\ \bibnamefont {Simien}}, \bibinfo
  {author} {\bibfnamefont {H.}~\bibnamefont {Gao}}, \bibinfo {author}
  {\bibfnamefont {J.}~\bibnamefont {Castro}}, \bibinfo {author} {\bibfnamefont
  {T.~C.}\ \bibnamefont {Killian}}, \ and\ \bibinfo {author} {\bibfnamefont
  {T.}~\bibnamefont {Pohl}},\ }\href@noop {} {\bibfield  {journal} {\bibinfo
  {journal} {{Phys. Rev. Lett.}}\ }\textbf {\bibinfo {volume} {99}},\ \bibinfo
  {pages} {75005} (\bibinfo {year} {2007})}\BibitemShut {NoStop}%
\bibitem [{\citenamefont {Roberts}\ \emph {et~al.}(2004)\citenamefont
  {Roberts}, \citenamefont {Fertig}, \citenamefont {Lim},\ and\ \citenamefont
  {Rolston}}]{rfl04}%
  \BibitemOpen
  \bibfield  {author} {\bibinfo {author} {\bibfnamefont {J.~L.}\ \bibnamefont
  {Roberts}}, \bibinfo {author} {\bibfnamefont {C.~D.}\ \bibnamefont {Fertig}},
  \bibinfo {author} {\bibfnamefont {M.~J.}\ \bibnamefont {Lim}}, \ and\
  \bibinfo {author} {\bibfnamefont {S.~L.}\ \bibnamefont {Rolston}},\ }\href
  {\doibase 10.1103/PhysRevLett.92.253003} {\bibfield  {journal} {\bibinfo
  {journal} {Phys. Rev. Lett.}\ }\textbf {\bibinfo {volume} {92}},\ \bibinfo
  {pages} {253003} (\bibinfo {year} {2004})}\BibitemShut {NoStop}%
\bibitem [{\citenamefont {Fletcher}, \citenamefont {Zhang},\ and\ \citenamefont
  {Rolston}(2007)}]{fzr07}%
  \BibitemOpen
  \bibfield  {author} {\bibinfo {author} {\bibfnamefont {R.~S.}\ \bibnamefont
  {Fletcher}}, \bibinfo {author} {\bibfnamefont {X.~L.}\ \bibnamefont {Zhang}},
  \ and\ \bibinfo {author} {\bibfnamefont {S.~L.}\ \bibnamefont {Rolston}},\
  }\href {\doibase 10.1103/PhysRevLett.99.145001} {\bibfield  {journal}
  {\bibinfo  {journal} {Phys. Rev. Lett.}\ }\textbf {\bibinfo {volume} {99}},\
  \bibinfo {eid} {145001} (\bibinfo {year} {2007})}\BibitemShut {NoStop}%
\bibitem [{\citenamefont {Pohl}, \citenamefont {Pattard},\ and\ \citenamefont
  {Rost}(2004{\natexlab{a}})}]{ppr04}%
  \BibitemOpen
  \bibfield  {author} {\bibinfo {author} {\bibfnamefont {T.}~\bibnamefont
  {Pohl}}, \bibinfo {author} {\bibfnamefont {T.}~\bibnamefont {Pattard}}, \
  and\ \bibinfo {author} {\bibfnamefont {J.~M.}\ \bibnamefont {Rost}},\
  }\href@noop {} {\bibfield  {journal} {\bibinfo  {journal} {{Phys. Rev.
  Lett.}}\ }\textbf {\bibinfo {volume} {92}},\ \bibinfo {pages} {155003}
  (\bibinfo {year} {2004}{\natexlab{a}})}\BibitemShut {NoStop}%
\bibitem [{\citenamefont {Pohl}, \citenamefont {Pattard},\ and\ \citenamefont
  {Rost}(2005{\natexlab{a}})}]{ppr05PRL}%
  \BibitemOpen
  \bibfield  {author} {\bibinfo {author} {\bibfnamefont {T.}~\bibnamefont
  {Pohl}}, \bibinfo {author} {\bibfnamefont {T.}~\bibnamefont {Pattard}}, \
  and\ \bibinfo {author} {\bibfnamefont {J.~M.}\ \bibnamefont {Rost}},\ }\href
  {\doibase 10.1103/PhysRevLett.94.205003} {\bibfield  {journal} {\bibinfo
  {journal} {Phys. Rev. Lett.}\ }\textbf {\bibinfo {volume} {94}},\ \bibinfo
  {pages} {205003} (\bibinfo {year} {2005}{\natexlab{a}})}\BibitemShut
  {NoStop}%
\bibitem [{\citenamefont {Li}\ \emph {et~al.}(1991)\citenamefont {Li},
  \citenamefont {Poggiani}, \citenamefont {Testera},\ and\ \citenamefont
  {Werth}}]{lpt91}%
  \BibitemOpen
  \bibfield  {author} {\bibinfo {author} {\bibfnamefont {G.}~\bibnamefont
  {Li}}, \bibinfo {author} {\bibfnamefont {R.}~\bibnamefont {Poggiani}},
  \bibinfo {author} {\bibfnamefont {G.}~\bibnamefont {Testera}}, \ and\
  \bibinfo {author} {\bibfnamefont {G.}~\bibnamefont {Werth}},\ }\href
  {\doibase 10.1007/BF01438559} {\bibfield  {journal} {\bibinfo  {journal}
  {Zeitschrift für Physik D Atoms, Molecules and Clusters}\ }\textbf {\bibinfo
  {volume} {22}},\ \bibinfo {pages} {375} (\bibinfo {year} {1991})}\BibitemShut
  {NoStop}%
\bibitem [{\citenamefont {Gabrielse}\ \emph {et~al.}(2011)\citenamefont
  {Gabrielse}, \citenamefont {Kolthammer}, \citenamefont {McConnell},
  \citenamefont {Richerme}, \citenamefont {Kalra}, \citenamefont {Novitski},
  \citenamefont {Grzonka}, \citenamefont {Oelert}, \citenamefont {Sefzick},
  \citenamefont {Zielinski}, \citenamefont {Fitzakerley}, \citenamefont
  {George}, \citenamefont {Hessels}, \citenamefont {Storry}, \citenamefont
  {Weel}, \citenamefont {M\"ullers},\ and\ \citenamefont {Walz}}]{gkm11}%
  \BibitemOpen
  \bibfield  {author} {\bibinfo {author} {\bibfnamefont {G.}~\bibnamefont
  {Gabrielse}}, \bibinfo {author} {\bibfnamefont {W.~S.}\ \bibnamefont
  {Kolthammer}}, \bibinfo {author} {\bibfnamefont {R.}~\bibnamefont
  {McConnell}}, \bibinfo {author} {\bibfnamefont {P.}~\bibnamefont {Richerme}},
  \bibinfo {author} {\bibfnamefont {R.}~\bibnamefont {Kalra}}, \bibinfo
  {author} {\bibfnamefont {E.}~\bibnamefont {Novitski}}, \bibinfo {author}
  {\bibfnamefont {D.}~\bibnamefont {Grzonka}}, \bibinfo {author} {\bibfnamefont
  {W.}~\bibnamefont {Oelert}}, \bibinfo {author} {\bibfnamefont
  {T.}~\bibnamefont {Sefzick}}, \bibinfo {author} {\bibfnamefont
  {M.}~\bibnamefont {Zielinski}}, \bibinfo {author} {\bibfnamefont
  {D.}~\bibnamefont {Fitzakerley}}, \bibinfo {author} {\bibfnamefont {M.~C.}\
  \bibnamefont {George}}, \bibinfo {author} {\bibfnamefont {E.~A.}\
  \bibnamefont {Hessels}}, \bibinfo {author} {\bibfnamefont {C.~H.}\
  \bibnamefont {Storry}}, \bibinfo {author} {\bibfnamefont {M.}~\bibnamefont
  {Weel}}, \bibinfo {author} {\bibfnamefont {A.}~\bibnamefont {M\"ullers}}, \
  and\ \bibinfo {author} {\bibfnamefont {J.}~\bibnamefont {Walz}} (\bibinfo
  {collaboration} {ATRAP Collaboration}),\ }\href {\doibase
  10.1103/PhysRevLett.106.073002} {\bibfield  {journal} {\bibinfo  {journal}
  {Phys. Rev. Lett.}\ }\textbf {\bibinfo {volume} {106}},\ \bibinfo {pages}
  {073002} (\bibinfo {year} {2011})}\BibitemShut {NoStop}%
\bibitem [{\citenamefont {Manfredi}\ and\ \citenamefont
  {Hervieux}(2012)}]{mh12}%
  \BibitemOpen
  \bibfield  {author} {\bibinfo {author} {\bibfnamefont {G.}~\bibnamefont
  {Manfredi}}\ and\ \bibinfo {author} {\bibfnamefont {P.-A.}\ \bibnamefont
  {Hervieux}},\ }\href {\doibase 10.1103/PhysRevLett.109.255005} {\bibfield
  {journal} {\bibinfo  {journal} {Phys. Rev. Lett.}\ }\textbf {\bibinfo
  {volume} {109}},\ \bibinfo {pages} {255005} (\bibinfo {year}
  {2012})}\BibitemShut {NoStop}%
\bibitem [{\citenamefont {Saul}, \citenamefont {Wurz},\ and\ \citenamefont
  {Kallenbach}(2009)}]{swk09}%
  \BibitemOpen
  \bibfield  {author} {\bibinfo {author} {\bibfnamefont {L.}~\bibnamefont
  {Saul}}, \bibinfo {author} {\bibfnamefont {P.}~\bibnamefont {Wurz}}, \ and\
  \bibinfo {author} {\bibfnamefont {R.}~\bibnamefont {Kallenbach}},\ }\href
  {http://stacks.iop.org/0004-637X/703/i=1/a=325} {\bibfield  {journal}
  {\bibinfo  {journal} {The Astrophysical Journal}\ }\textbf {\bibinfo {volume}
  {703}},\ \bibinfo {pages} {325} (\bibinfo {year} {2009})}\BibitemShut
  {NoStop}%
\bibitem [{\citenamefont {Mark}\ \emph {et~al.}(2012)\citenamefont {Mark},
  \citenamefont {Haller}, \citenamefont {Lauber}, \citenamefont {Danzl},
  \citenamefont {Janisch}, \citenamefont {B\"{u}chler}, \citenamefont {Daley},\
  and\ \citenamefont {N\"{a}gerl}}]{mhl12}%
  \BibitemOpen
  \bibfield  {author} {\bibinfo {author} {\bibfnamefont {M.}~\bibnamefont
  {Mark}}, \bibinfo {author} {\bibfnamefont {E.}~\bibnamefont {Haller}},
  \bibinfo {author} {\bibfnamefont {K.}~\bibnamefont {Lauber}}, \bibinfo
  {author} {\bibfnamefont {J.}~\bibnamefont {Danzl}}, \bibinfo {author}
  {\bibfnamefont {A.}~\bibnamefont {Janisch}}, \bibinfo {author} {\bibfnamefont
  {H.}~\bibnamefont {B\"{u}chler}}, \bibinfo {author} {\bibfnamefont
  {A.}~\bibnamefont {Daley}}, \ and\ \bibinfo {author} {\bibfnamefont {H.-C.}\
  \bibnamefont {N\"{a}gerl}},\ }\href
  {http://prl.aps.org/abstract/PRL/v108/i21/e215302} {\bibfield  {journal}
  {\bibinfo  {journal} {Phys. Rev. Lett.}\ }\textbf {\bibinfo {volume} {108}},\
  \bibinfo {pages} {215302} (\bibinfo {year} {2012})}\BibitemShut {NoStop}%
\bibitem [{\citenamefont {Chen}\ \emph {et~al.}(2013)\citenamefont {Chen},
  \citenamefont {M\"obius}, \citenamefont {Gloeckler}, \citenamefont
  {Bochsler}, \citenamefont {Bzowski}, \citenamefont {Isenberg},\ and\
  \citenamefont {Sok\'o}}]{cmg13}%
  \BibitemOpen
  \bibfield  {author} {\bibinfo {author} {\bibfnamefont {J.~H.}\ \bibnamefont
  {Chen}}, \bibinfo {author} {\bibfnamefont {E.}~\bibnamefont {M\"obius}},
  \bibinfo {author} {\bibfnamefont {G.}~\bibnamefont {Gloeckler}}, \bibinfo
  {author} {\bibfnamefont {P.}~\bibnamefont {Bochsler}}, \bibinfo {author}
  {\bibfnamefont {M.}~\bibnamefont {Bzowski}}, \bibinfo {author} {\bibfnamefont
  {P.~A.}\ \bibnamefont {Isenberg}}, \ and\ \bibinfo {author} {\bibfnamefont
  {J.~M.}\ \bibnamefont {Sok\'o}},\ }\href {\doibase 10.1002/jgra.50391}
  {\bibfield  {journal} {\bibinfo  {journal} {Journal of Geophysical Research:
  Space Physics}\ }\textbf {\bibinfo {volume} {118}},\ \bibinfo {pages} {3946}
  (\bibinfo {year} {2013})}\BibitemShut {NoStop}%
\bibitem [{\citenamefont {Betti}\ \emph {et~al.}(2005)\citenamefont {Betti},
  \citenamefont {Ceccherini}, \citenamefont {Cornolti},\ and\ \citenamefont
  {Pegoraro}}]{bcc05}%
  \BibitemOpen
  \bibfield  {author} {\bibinfo {author} {\bibfnamefont {S.}~\bibnamefont
  {Betti}}, \bibinfo {author} {\bibfnamefont {F.}~\bibnamefont {Ceccherini}},
  \bibinfo {author} {\bibfnamefont {F.}~\bibnamefont {Cornolti}}, \ and\
  \bibinfo {author} {\bibfnamefont {F.}~\bibnamefont {Pegoraro}},\ }\href@noop
  {} {\bibfield  {journal} {\bibinfo  {journal} {Plasma Physics and Controlled
  Fusion}\ }\textbf {\bibinfo {volume} {47}},\ \bibinfo {pages} {521} (\bibinfo
  {year} {2005})}\BibitemShut {NoStop}%
\bibitem [{\citenamefont {Ceccherini}\ \emph {et~al.}(2006)\citenamefont
  {Ceccherini}, \citenamefont {Betti}, \citenamefont {Cornolti},\ and\
  \citenamefont {Pegoraro}}]{cbc06}%
  \BibitemOpen
  \bibfield  {author} {\bibinfo {author} {\bibfnamefont {F.}~\bibnamefont
  {Ceccherini}}, \bibinfo {author} {\bibfnamefont {S.}~\bibnamefont {Betti}},
  \bibinfo {author} {\bibfnamefont {F.}~\bibnamefont {Cornolti}}, \ and\
  \bibinfo {author} {\bibfnamefont {F.}~\bibnamefont {Pegoraro}},\ }\href
  {\doibase 10.1134/S1054660X06040104} {\bibfield  {journal} {\bibinfo
  {journal} {Laser Physics}\ }\textbf {\bibinfo {volume} {16}},\ \bibinfo
  {pages} {594} (\bibinfo {year} {2006})}\BibitemShut {NoStop}%
\bibitem [{\citenamefont {Castro}, \citenamefont {Gao},\ and\ \citenamefont
  {Killian}(2008)}]{cgk08}%
  \BibitemOpen
  \bibfield  {author} {\bibinfo {author} {\bibfnamefont {J.}~\bibnamefont
  {Castro}}, \bibinfo {author} {\bibfnamefont {H.}~\bibnamefont {Gao}}, \ and\
  \bibinfo {author} {\bibfnamefont {T.~C.}\ \bibnamefont {Killian}},\
  }\href@noop {} {\bibfield  {journal} {\bibinfo  {journal} {Plasma Physics of
  Controlled Fusion}\ }\textbf {\bibinfo {volume} {50}},\ \bibinfo {pages}
  {124011} (\bibinfo {year} {2008})}\BibitemShut {NoStop}%
\bibitem [{\citenamefont {Laha}\ \emph {et~al.}(2006)\citenamefont {Laha},
  \citenamefont {Chen}, \citenamefont {Gupta}, \citenamefont {Simien},
  \citenamefont {Martinez}, \citenamefont {Mickelson}, \citenamefont {Nagel},\
  and\ \citenamefont {Killian}}]{lcg06}%
  \BibitemOpen
  \bibfield  {author} {\bibinfo {author} {\bibfnamefont {S.}~\bibnamefont
  {Laha}}, \bibinfo {author} {\bibfnamefont {Y.~C.}\ \bibnamefont {Chen}},
  \bibinfo {author} {\bibfnamefont {P.}~\bibnamefont {Gupta}}, \bibinfo
  {author} {\bibfnamefont {C.~E.}\ \bibnamefont {Simien}}, \bibinfo {author}
  {\bibfnamefont {Y.~N.}\ \bibnamefont {Martinez}}, \bibinfo {author}
  {\bibfnamefont {P.~G.}\ \bibnamefont {Mickelson}}, \bibinfo {author}
  {\bibfnamefont {S.~B.}\ \bibnamefont {Nagel}}, \ and\ \bibinfo {author}
  {\bibfnamefont {T.~C.}\ \bibnamefont {Killian}},\ }\href@noop {} {\bibfield
  {journal} {\bibinfo  {journal} {{European Physical Journal D}}\ }\textbf
  {\bibinfo {volume} {40}},\ \bibinfo {pages} {51} (\bibinfo {year}
  {2006})}\BibitemShut {NoStop}%
\bibitem [{\citenamefont {McQuillen}(2012)}]{pcmThesis12}%
  \BibitemOpen
  \bibfield  {author} {\bibinfo {author} {\bibfnamefont {P.}~\bibnamefont
  {McQuillen}},\ }\emph {\bibinfo {title} {High Resolution Sculpting and
  Imaging of Ultracold Neutral Plasmas}},\ \href@noop {} {\bibinfo {type}
  {Master of science thesis}},\ \bibinfo  {school} {Rice University} (\bibinfo
  {year} {2012})\BibitemShut {NoStop}%
\bibitem [{\citenamefont {Castleman}(2007)}]{cas07}%
  \BibitemOpen
  \bibfield  {author} {\bibinfo {author} {\bibfnamefont {K.~R.}\ \bibnamefont
  {Castleman}},\ }\href {http://books.google.com/books?id=w5PGKs2TwiIC} {\emph
  {\bibinfo {title} {Digital Image Processing}}}\ (\bibinfo  {publisher}
  {Pearson Education},\ \bibinfo {year} {2007})\BibitemShut {NoStop}%
\bibitem [{\citenamefont {Simien}\ \emph {et~al.}(2004)\citenamefont {Simien},
  \citenamefont {Chen}, \citenamefont {Gupta}, \citenamefont {Laha},
  \citenamefont {Martinez}, \citenamefont {Mickelson}, \citenamefont {Nagel},\
  and\ \citenamefont {Killian}}]{scg04}%
  \BibitemOpen
  \bibfield  {author} {\bibinfo {author} {\bibfnamefont {C.~E.}\ \bibnamefont
  {Simien}}, \bibinfo {author} {\bibfnamefont {Y.~C.}\ \bibnamefont {Chen}},
  \bibinfo {author} {\bibfnamefont {P.}~\bibnamefont {Gupta}}, \bibinfo
  {author} {\bibfnamefont {S.}~\bibnamefont {Laha}}, \bibinfo {author}
  {\bibfnamefont {Y.~N.}\ \bibnamefont {Martinez}}, \bibinfo {author}
  {\bibfnamefont {P.~G.}\ \bibnamefont {Mickelson}}, \bibinfo {author}
  {\bibfnamefont {S.~B.}\ \bibnamefont {Nagel}}, \ and\ \bibinfo {author}
  {\bibfnamefont {T.~C.}\ \bibnamefont {Killian}},\ }\href@noop {} {\bibfield
  {journal} {\bibinfo  {journal} {{Phys. Rev. Lett.}}\ }\textbf {\bibinfo
  {volume} {92}},\ \bibinfo {pages} {143001} (\bibinfo {year}
  {2004})}\BibitemShut {NoStop}%
\bibitem [{\citenamefont {Garton}\ and\ \citenamefont {Codling}(1968)}]{gc68}%
  \BibitemOpen
  \bibfield  {author} {\bibinfo {author} {\bibfnamefont {W.~R.~S.}\
  \bibnamefont {Garton}}\ and\ \bibinfo {author} {\bibfnamefont
  {K.}~\bibnamefont {Codling}},\ }\href
  {http://stacks.iop.org/0022-3700/1/i=1/a=316} {\bibfield  {journal} {\bibinfo
   {journal} {Journal of Physics B: Atomic and Molecular Physics}\ }\textbf
  {\bibinfo {volume} {1}},\ \bibinfo {pages} {106} (\bibinfo {year}
  {1968})}\BibitemShut {NoStop}%
\bibitem [{\citenamefont {Killian}\ \emph {et~al.}(1999)\citenamefont
  {Killian}, \citenamefont {Kulin}, \citenamefont {Bergeson}, \citenamefont
  {Orozco}, \citenamefont {Orzel},\ and\ \citenamefont {Rolston}}]{kkb99}%
  \BibitemOpen
  \bibfield  {author} {\bibinfo {author} {\bibfnamefont {T.~C.}\ \bibnamefont
  {Killian}}, \bibinfo {author} {\bibfnamefont {S.}~\bibnamefont {Kulin}},
  \bibinfo {author} {\bibfnamefont {S.~D.}\ \bibnamefont {Bergeson}}, \bibinfo
  {author} {\bibfnamefont {L.~A.}\ \bibnamefont {Orozco}}, \bibinfo {author}
  {\bibfnamefont {C.}~\bibnamefont {Orzel}}, \ and\ \bibinfo {author}
  {\bibfnamefont {S.~L.}\ \bibnamefont {Rolston}},\ }\href@noop {} {\bibfield
  {journal} {\bibinfo  {journal} {{Phys. Rev. Lett.}}\ }\textbf {\bibinfo
  {volume} {83}},\ \bibinfo {pages} {4776} (\bibinfo {year}
  {1999})}\BibitemShut {NoStop}%
\bibitem [{\citenamefont {Donko}\ and\ \citenamefont {Hartmann}(2004)}]{dha04}%
  \BibitemOpen
  \bibfield  {author} {\bibinfo {author} {\bibfnamefont {Z.}~\bibnamefont
  {Donko}}\ and\ \bibinfo {author} {\bibfnamefont {P.}~\bibnamefont
  {Hartmann}},\ }\href@noop {} {\bibfield  {journal} {\bibinfo  {journal}
  {Phys. Rev. E}\ }\textbf {\bibinfo {volume} {69}},\ \bibinfo {eid} {016405}
  (\bibinfo {year} {2004})}\BibitemShut {NoStop}%
\bibitem [{\citenamefont {Hamaguchi}, \citenamefont {Farouki},\ and\
  \citenamefont {Dubin}(1997)}]{hfd97}%
  \BibitemOpen
  \bibfield  {author} {\bibinfo {author} {\bibfnamefont {S.}~\bibnamefont
  {Hamaguchi}}, \bibinfo {author} {\bibfnamefont {R.~T.}\ \bibnamefont
  {Farouki}}, \ and\ \bibinfo {author} {\bibfnamefont {D.~H.~E.}\ \bibnamefont
  {Dubin}},\ }\href@noop {} {\bibfield  {journal} {\bibinfo  {journal} {{Phys.
  Rev. E}}\ }\textbf {\bibinfo {volume} {56}},\ \bibinfo {pages} {4671}
  (\bibinfo {year} {1997})}\BibitemShut {NoStop}%
\bibitem [{\citenamefont {Guo}, \citenamefont {Lu},\ and\ \citenamefont
  {Han}(2010)}]{glh10}%
  \BibitemOpen
  \bibfield  {author} {\bibinfo {author} {\bibfnamefont {L.}~\bibnamefont
  {Guo}}, \bibinfo {author} {\bibfnamefont {R.~H.}\ \bibnamefont {Lu}}, \ and\
  \bibinfo {author} {\bibfnamefont {S.~S.}\ \bibnamefont {Han}},\ }\href@noop
  {} {\bibfield  {journal} {\bibinfo  {journal} {Phys. Rev. E}\ }\textbf
  {\bibinfo {volume} {81}},\ \bibinfo {pages} {046406} (\bibinfo {year}
  {2010})}\BibitemShut {NoStop}%
\bibitem [{\citenamefont {Lyon}\ and\ \citenamefont {Bergeson}(2011)}]{lb11}%
  \BibitemOpen
  \bibfield  {author} {\bibinfo {author} {\bibfnamefont {M.}~\bibnamefont
  {Lyon}}\ and\ \bibinfo {author} {\bibfnamefont {S.~D.}\ \bibnamefont
  {Bergeson}},\ }\href {http://stacks.iop.org/0953-4075/44/i=18/a=184014}
  {\bibfield  {journal} {\bibinfo  {journal} {Journal of Physics B: Atomic,
  Molecular and Optical Physics}\ }\textbf {\bibinfo {volume} {44}},\ \bibinfo
  {pages} {184014} (\bibinfo {year} {2011})}\BibitemShut {NoStop}%
\bibitem [{poh()}]{pohlPrv15}%
  \BibitemOpen
  \href@noop {} {}\bibinfo {note} {T. Pohl, Private communication
  (2014).}\BibitemShut {Stop}%
\bibitem [{\citenamefont {Langin}\ \emph {et~al.}()\citenamefont {Langin},
  \citenamefont {McQuillen}, \citenamefont {Strickler},\ and\ \citenamefont
  {Killian}}]{tbp15}%
  \BibitemOpen
  \bibfield  {author} {\bibinfo {author} {\bibfnamefont {T.}~\bibnamefont
  {Langin}}, \bibinfo {author} {\bibfnamefont {P.}~\bibnamefont {McQuillen}},
  \bibinfo {author} {\bibfnamefont {T.}~\bibnamefont {Strickler}}, \ and\
  \bibinfo {author} {\bibfnamefont {T.}~\bibnamefont {Killian}},\ }\href@noop
  {} {\bibinfo  {journal} {To be published}\ }\BibitemShut {NoStop}%
\bibitem [{\citenamefont {Pohl}, \citenamefont {Pattard},\ and\ \citenamefont
  {Rost}(2004{\natexlab{b}})}]{ppr04PRA}%
  \BibitemOpen
\bibfield  {journal} {  }\bibfield  {author} {\bibinfo {author} {\bibfnamefont
  {T.}~\bibnamefont {Pohl}}, \bibinfo {author} {\bibfnamefont {T.}~\bibnamefont
  {Pattard}}, \ and\ \bibinfo {author} {\bibfnamefont {J.~M.}\ \bibnamefont
  {Rost}},\ }\href@noop {} {\bibfield  {journal} {\bibinfo  {journal} {{Phys.
  Rev. A}}\ }\textbf {\bibinfo {volume} {70}},\ \bibinfo {pages} {033416}
  (\bibinfo {year} {2004}{\natexlab{b}})}\BibitemShut {NoStop}%
\bibitem [{\citenamefont {Chabrier}\ and\ \citenamefont
  {Potekhin}(1998)}]{cp98}%
  \BibitemOpen
  \bibfield  {author} {\bibinfo {author} {\bibfnamefont {G.}~\bibnamefont
  {Chabrier}}\ and\ \bibinfo {author} {\bibfnamefont {A.~Y.}\ \bibnamefont
  {Potekhin}},\ }\href {\doibase 10.1103/PhysRevE.58.4941} {\bibfield
  {journal} {\bibinfo  {journal} {Phys. Rev. E}\ }\textbf {\bibinfo {volume}
  {58}},\ \bibinfo {pages} {4941} (\bibinfo {year} {1998})}\BibitemShut
  {NoStop}%
\bibitem [{\citenamefont {Kuzmin}\ and\ \citenamefont
  {O'Neil}(2002{\natexlab{b}})}]{kon02}%
  \BibitemOpen
  \bibfield  {author} {\bibinfo {author} {\bibfnamefont {S.~G.}\ \bibnamefont
  {Kuzmin}}\ and\ \bibinfo {author} {\bibfnamefont {T.~M.}\ \bibnamefont
  {O'Neil}},\ }\href@noop {} {\bibfield  {journal} {\bibinfo  {journal} {{Phys.
  Plasmas}}\ }\textbf {\bibinfo {volume} {9}},\ \bibinfo {pages} {3743}
  (\bibinfo {year} {2002}{\natexlab{b}})}\BibitemShut {NoStop}%
\bibitem [{\citenamefont {Pohl}, \citenamefont {Pattard},\ and\ \citenamefont
  {Rost}(2005{\natexlab{b}})}]{ppr05JPB}%
  \BibitemOpen
  \bibfield  {author} {\bibinfo {author} {\bibfnamefont {T.}~\bibnamefont
  {Pohl}}, \bibinfo {author} {\bibfnamefont {T.}~\bibnamefont {Pattard}}, \
  and\ \bibinfo {author} {\bibfnamefont {J.}~\bibnamefont {Rost}},\ }\href@noop
  {} {\bibfield  {journal} {\bibinfo  {journal} {J.\ Phys.\ B}\ }\textbf
  {\bibinfo {volume} {38}},\ \bibinfo {pages} {S343} (\bibinfo {year}
  {2005}{\natexlab{b}})}\BibitemShut {NoStop}%
\bibitem [{\citenamefont {Dawson}(1964)}]{daw64}%
  \BibitemOpen
  \bibfield  {author} {\bibinfo {author} {\bibfnamefont {J.~M.}\ \bibnamefont
  {Dawson}},\ }\href {\doibase http://dx.doi.org/10.1063/1.1711214} {\bibfield
  {journal} {\bibinfo  {journal} {Physics of Fluids}\ }\textbf {\bibinfo
  {volume} {7}},\ \bibinfo {pages} {419} (\bibinfo {year} {1964})}\BibitemShut
  {NoStop}%
\bibitem [{\citenamefont {Montgomery}\ and\ \citenamefont
  {Nielson}(1970)}]{mn70}%
  \BibitemOpen
  \bibfield  {author} {\bibinfo {author} {\bibfnamefont {D.}~\bibnamefont
  {Montgomery}}\ and\ \bibinfo {author} {\bibfnamefont {C.~W.}\ \bibnamefont
  {Nielson}},\ }\href {\doibase http://dx.doi.org/10.1063/1.1693081} {\bibfield
   {journal} {\bibinfo  {journal} {Physics of Fluids}\ }\textbf {\bibinfo
  {volume} {13}},\ \bibinfo {pages} {1405} (\bibinfo {year}
  {1970})}\BibitemShut {NoStop}%
\bibitem [{\citenamefont {Boercker}\ and\ \citenamefont {More}(1986)}]{bm86}%
  \BibitemOpen
  \bibfield  {author} {\bibinfo {author} {\bibfnamefont {D.~B.}\ \bibnamefont
  {Boercker}}\ and\ \bibinfo {author} {\bibfnamefont {R.~M.}\ \bibnamefont
  {More}},\ }\href {\doibase 10.1103/PhysRevA.33.1859} {\bibfield  {journal}
  {\bibinfo  {journal} {Phys. Rev. A}\ }\textbf {\bibinfo {volume} {33}},\
  \bibinfo {pages} {1859} (\bibinfo {year} {1986})}\BibitemShut {NoStop}%
\bibitem [{\citenamefont {Hyatt}, \citenamefont {Driscoll},\ and\ \citenamefont
  {Malmberg}(1987)}]{hdm87}%
  \BibitemOpen
  \bibfield  {author} {\bibinfo {author} {\bibfnamefont {A.~W.}\ \bibnamefont
  {Hyatt}}, \bibinfo {author} {\bibfnamefont {C.~F.}\ \bibnamefont {Driscoll}},
  \ and\ \bibinfo {author} {\bibfnamefont {J.~H.}\ \bibnamefont {Malmberg}},\
  }\href {\doibase 10.1103/PhysRevLett.59.2975} {\bibfield  {journal} {\bibinfo
   {journal} {Phys. Rev. Lett.}\ }\textbf {\bibinfo {volume} {59}},\ \bibinfo
  {pages} {2975} (\bibinfo {year} {1987})}\BibitemShut {NoStop}%
\bibitem [{\citenamefont {Martin}(1999)}]{mar99}%
  \BibitemOpen
  \bibfield  {author} {\bibinfo {author} {\bibfnamefont {P.}~\bibnamefont
  {Martin}},\ }\href {http://stacks.iop.org/0741-3335/41/i=3A/a=018} {\bibfield
   {journal} {\bibinfo  {journal} {Plasma Physics and Controlled Fusion}\
  }\textbf {\bibinfo {volume} {41}},\ \bibinfo {pages} {A247} (\bibinfo {year}
  {1999})}\BibitemShut {NoStop}%
\bibitem [{\citenamefont {Morozov}\ and\ \citenamefont {Norman}(2003)}]{mn03}%
  \BibitemOpen
  \bibfield  {author} {\bibinfo {author} {\bibfnamefont {I.~V.}\ \bibnamefont
  {Morozov}}\ and\ \bibinfo {author} {\bibfnamefont {G.~E.}\ \bibnamefont
  {Norman}},\ }\href {http://stacks.iop.org/0305-4470/36/i=22/a=323} {\bibfield
   {journal} {\bibinfo  {journal} {Journal of Physics A: Mathematical and
  General}\ }\textbf {\bibinfo {volume} {36}},\ \bibinfo {pages} {6005}
  (\bibinfo {year} {2003})}\BibitemShut {NoStop}%
\bibitem [{\citenamefont {Beck}, \citenamefont {Fajans},\ and\ \citenamefont
  {Malmberg}(1996)}]{bfm06}%
  \BibitemOpen
  \bibfield  {author} {\bibinfo {author} {\bibfnamefont {B.~R.}\ \bibnamefont
  {Beck}}, \bibinfo {author} {\bibfnamefont {J.}~\bibnamefont {Fajans}}, \ and\
  \bibinfo {author} {\bibfnamefont {J.~H.}\ \bibnamefont {Malmberg}},\ }\href
  {\doibase http://dx.doi.org/10.1063/1.871749} {\bibfield  {journal} {\bibinfo
   {journal} {Phys. Plasmas}\ }\textbf {\bibinfo {volume} {3}},\ \bibinfo
  {pages} {1250} (\bibinfo {year} {1996})}\BibitemShut {NoStop}%
\bibitem [{\citenamefont {Dimonte}\ and\ \citenamefont
  {Daligault}(2008)}]{dd08}%
  \BibitemOpen
  \bibfield  {author} {\bibinfo {author} {\bibfnamefont {G.}~\bibnamefont
  {Dimonte}}\ and\ \bibinfo {author} {\bibfnamefont {J.}~\bibnamefont
  {Daligault}},\ }\href {\doibase 10.1103/PhysRevLett.101.135001} {\bibfield
  {journal} {\bibinfo  {journal} {Phys. Rev. Lett.}\ }\textbf {\bibinfo
  {volume} {101}},\ \bibinfo {pages} {135001} (\bibinfo {year}
  {2008})}\BibitemShut {NoStop}%
\bibitem [{\citenamefont {Murillo}\ and\ \citenamefont
  {Dharma-Wardana}(2008)}]{mdh08}%
  \BibitemOpen
  \bibfield  {author} {\bibinfo {author} {\bibfnamefont {M.~S.}\ \bibnamefont
  {Murillo}}\ and\ \bibinfo {author} {\bibfnamefont {M.~W.~C.}\ \bibnamefont
  {Dharma-Wardana}},\ }\href
  {http://prl.aps.org.ezproxy.rice.edu/abstract/PRL/v100/i20/e205005}
  {\bibfield  {journal} {\bibinfo  {journal} {Phys. Rev. Lett.}\ }\textbf
  {\bibinfo {volume} {100}},\ \bibinfo {pages} {205005} (\bibinfo {year}
  {2008})}\BibitemShut {NoStop}%
\bibitem [{\citenamefont {Glosli}\ \emph {et~al.}(2008)\citenamefont {Glosli},
  \citenamefont {Graziani}, \citenamefont {More}, \citenamefont {Murillo},
  \citenamefont {Streitz}, \citenamefont {Surh}, \citenamefont {Benedict},
  \citenamefont {Hau-Riege}, \citenamefont {Langdon},\ and\ \citenamefont
  {London}}]{ggm08}%
  \BibitemOpen
  \bibfield  {author} {\bibinfo {author} {\bibfnamefont {J.~N.}\ \bibnamefont
  {Glosli}}, \bibinfo {author} {\bibfnamefont {F.~R.}\ \bibnamefont
  {Graziani}}, \bibinfo {author} {\bibfnamefont {R.~M.}\ \bibnamefont {More}},
  \bibinfo {author} {\bibfnamefont {M.~S.}\ \bibnamefont {Murillo}}, \bibinfo
  {author} {\bibfnamefont {F.~H.}\ \bibnamefont {Streitz}}, \bibinfo {author}
  {\bibfnamefont {M.~P.}\ \bibnamefont {Surh}}, \bibinfo {author}
  {\bibfnamefont {L.~X.}\ \bibnamefont {Benedict}}, \bibinfo {author}
  {\bibfnamefont {S.}~\bibnamefont {Hau-Riege}}, \bibinfo {author}
  {\bibfnamefont {A.~B.}\ \bibnamefont {Langdon}}, \ and\ \bibinfo {author}
  {\bibfnamefont {R.~A.}\ \bibnamefont {London}},\ }\href@noop {} {\bibfield
  {journal} {\bibinfo  {journal} {Phys. Rev. E}\ }\textbf {\bibinfo {volume}
  {78}},\ \bibinfo {pages} {025401R} (\bibinfo {year} {2008})}\BibitemShut
  {NoStop}%
\bibitem [{\citenamefont {Rygg}\ \emph {et~al.}(2009)\citenamefont {Rygg},
  \citenamefont {Frenje}, \citenamefont {Li}, \citenamefont {S\'eguin},
  \citenamefont {Petrasso}, \citenamefont {Meyerhofer},\ and\ \citenamefont
  {Stoeckl}}]{rfl09}%
  \BibitemOpen
  \bibfield  {author} {\bibinfo {author} {\bibfnamefont {J.~R.}\ \bibnamefont
  {Rygg}}, \bibinfo {author} {\bibfnamefont {J.~A.}\ \bibnamefont {Frenje}},
  \bibinfo {author} {\bibfnamefont {C.~K.}\ \bibnamefont {Li}}, \bibinfo
  {author} {\bibfnamefont {F.~H.}\ \bibnamefont {S\'eguin}}, \bibinfo {author}
  {\bibfnamefont {R.~D.}\ \bibnamefont {Petrasso}}, \bibinfo {author}
  {\bibfnamefont {D.~D.}\ \bibnamefont {Meyerhofer}}, \ and\ \bibinfo {author}
  {\bibfnamefont {C.}~\bibnamefont {Stoeckl}},\ }\href
  {http://link.aps.org/doi/10.1103/PhysRevE.80.026403} {\bibfield  {journal}
  {\bibinfo  {journal} {Phys. Rev. E}\ }\textbf {\bibinfo {volume} {80}},\
  \bibinfo {pages} {026403} (\bibinfo {year} {2009})}\BibitemShut {NoStop}%
\bibitem [{\citenamefont {Vorberger}\ \emph {et~al.}(2010)\citenamefont
  {Vorberger}, \citenamefont {Gericke}, \citenamefont {Bornath},\ and\
  \citenamefont {Schlanges}}]{vgb10}%
  \BibitemOpen
  \bibfield  {author} {\bibinfo {author} {\bibfnamefont {J.}~\bibnamefont
  {Vorberger}}, \bibinfo {author} {\bibfnamefont {D.~O.}\ \bibnamefont
  {Gericke}}, \bibinfo {author} {\bibfnamefont {T.}~\bibnamefont {Bornath}}, \
  and\ \bibinfo {author} {\bibfnamefont {M.}~\bibnamefont {Schlanges}},\
  }\href@noop {} {\bibfield  {journal} {\bibinfo  {journal} {Phys. Rev. E}\
  }\textbf {\bibinfo {volume} {81}},\ \bibinfo {pages} {046404} (\bibinfo
  {year} {2010})}\BibitemShut {NoStop}%
\bibitem [{\citenamefont {Benedict}\ \emph {et~al.}(2012)\citenamefont
  {Benedict}, \citenamefont {Surh}, \citenamefont {Castor}, \citenamefont
  {Khairallah}, \citenamefont {Whitley}, \citenamefont {Richards},
  \citenamefont {Glosli}, \citenamefont {Murillo}, \citenamefont {Scullard},
  \citenamefont {Grabowski}, \citenamefont {Michta},\ and\ \citenamefont
  {Graziani}}]{bsc12}%
  \BibitemOpen
  \bibfield  {author} {\bibinfo {author} {\bibfnamefont {L.~X.}\ \bibnamefont
  {Benedict}}, \bibinfo {author} {\bibfnamefont {M.~P.}\ \bibnamefont {Surh}},
  \bibinfo {author} {\bibfnamefont {J.~I.}\ \bibnamefont {Castor}}, \bibinfo
  {author} {\bibfnamefont {S.~A.}\ \bibnamefont {Khairallah}}, \bibinfo
  {author} {\bibfnamefont {H.~D.}\ \bibnamefont {Whitley}}, \bibinfo {author}
  {\bibfnamefont {D.~F.}\ \bibnamefont {Richards}}, \bibinfo {author}
  {\bibfnamefont {J.~N.}\ \bibnamefont {Glosli}}, \bibinfo {author}
  {\bibfnamefont {M.~S.}\ \bibnamefont {Murillo}}, \bibinfo {author}
  {\bibfnamefont {C.~R.}\ \bibnamefont {Scullard}}, \bibinfo {author}
  {\bibfnamefont {P.~E.}\ \bibnamefont {Grabowski}}, \bibinfo {author}
  {\bibfnamefont {D.}~\bibnamefont {Michta}}, \ and\ \bibinfo {author}
  {\bibfnamefont {F.~R.}\ \bibnamefont {Graziani}},\ }\href@noop {} {\bibfield
  {journal} {\bibinfo  {journal} {Phys. Rev. E}\ }\textbf {\bibinfo {volume}
  {86}},\ \bibinfo {pages} {046406} (\bibinfo {year} {2012})}\BibitemShut
  {NoStop}%
\bibitem [{\citenamefont {Schlanges}\ \emph {et~al.}(2010)\citenamefont
  {Schlanges}, \citenamefont {Bornath}, \citenamefont {Vorberger},\ and\
  \citenamefont {Gericke}}]{sbv10}%
  \BibitemOpen
  \bibfield  {author} {\bibinfo {author} {\bibfnamefont {M.}~\bibnamefont
  {Schlanges}}, \bibinfo {author} {\bibfnamefont {T.}~\bibnamefont {Bornath}},
  \bibinfo {author} {\bibfnamefont {J.}~\bibnamefont {Vorberger}}, \ and\
  \bibinfo {author} {\bibfnamefont {D.}~\bibnamefont {Gericke}},\ }\href
  {\doibase 10.1002/ctpp.201010014} {\bibfield  {journal} {\bibinfo  {journal}
  {Contributions to Plasma Physics}\ }\textbf {\bibinfo {volume} {50}},\
  \bibinfo {pages} {64} (\bibinfo {year} {2010})}\BibitemShut {NoStop}%
\bibitem [{\citenamefont {Papadopoulos}(1971)}]{pap71}%
  \BibitemOpen
  \bibfield  {author} {\bibinfo {author} {\bibfnamefont {K.}~\bibnamefont
  {Papadopoulos}},\ }\href {\doibase 10.1029/JA076i016p03806} {\bibfield
  {journal} {\bibinfo  {journal} {Journal of Geophysical Research}\ }\textbf
  {\bibinfo {volume} {76}},\ \bibinfo {pages} {3806} (\bibinfo {year}
  {1971})}\BibitemShut {NoStop}%
\bibitem [{\citenamefont {Marsch}(1991)}]{sm91}%
  \BibitemOpen
  \bibfield  {author} {\bibinfo {author} {\bibfnamefont {E.}~\bibnamefont
  {Marsch}},\ }in\ \href {\doibase 10.1007/978-3-642-75364-0_3} {\emph
  {\bibinfo {booktitle} {Physics of the Inner Heliosphere II}}},\ \bibinfo
  {series} {Physics and Chemistry in Space}, Vol.~\bibinfo {volume} {21},\
  \bibinfo {editor} {edited by\ \bibinfo {editor} {\bibfnamefont
  {R.}~\bibnamefont {Schwenn}}\ and\ \bibinfo {editor} {\bibfnamefont
  {E.}~\bibnamefont {Marsch}}}\ (\bibinfo  {publisher} {Springer Berlin
  Heidelberg},\ \bibinfo {year} {1991})\ pp.\ \bibinfo {pages}
  {45--133}\BibitemShut {NoStop}%
\bibitem [{\citenamefont {Gnedin}, \citenamefont {Yakovlev},\ and\
  \citenamefont {Potekhin}(2001)}]{gyp01}%
  \BibitemOpen
  \bibfield  {author} {\bibinfo {author} {\bibfnamefont {O.~Y.}\ \bibnamefont
  {Gnedin}}, \bibinfo {author} {\bibfnamefont {D.~G.}\ \bibnamefont
  {Yakovlev}}, \ and\ \bibinfo {author} {\bibfnamefont {A.~Y.}\ \bibnamefont
  {Potekhin}},\ }\href {\doibase 10.1046/j.1365-8711.2001.04359.x} {\bibfield
  {journal} {\bibinfo  {journal} {Monthly Notices of the Royal Astronomical
  Society}\ }\textbf {\bibinfo {volume} {324}},\ \bibinfo {pages} {725}
  (\bibinfo {year} {2001})}\BibitemShut {NoStop}%
\bibitem [{\citenamefont {Spitzer}(1956)}]{spi56}%
  \BibitemOpen
  \bibfield  {author} {\bibinfo {author} {\bibfnamefont {L.}~\bibnamefont
  {Spitzer}},\ }\href {http://books.google.com/books?id=Cb8EAAAAMAAJ} {\emph
  {\bibinfo {title} {Physics of Fully Ionized Gases}}},\ Interscience tracts on
  physics and astronomy, 3\ (\bibinfo  {publisher} {Interscience Publishers},\
  \bibinfo {year} {1956})\BibitemShut {NoStop}%
\bibitem [{\citenamefont {Hansen}\ and\ \citenamefont
  {McDonald}(1983)}]{hmc83}%
  \BibitemOpen
  \bibfield  {author} {\bibinfo {author} {\bibfnamefont {J.~P.}\ \bibnamefont
  {Hansen}}\ and\ \bibinfo {author} {\bibfnamefont {I.~R.}\ \bibnamefont
  {McDonald}},\ }\href@noop {} {\bibfield  {journal} {\bibinfo  {journal}
  {{Physics Letters A}}\ }\textbf {\bibinfo {volume} {97}},\ \bibinfo {pages}
  {42} (\bibinfo {year} {1983})}\BibitemShut {NoStop}%
\bibitem [{\citenamefont {Gericke}, \citenamefont {Murillo},\ and\
  \citenamefont {Schlanges}(2002)}]{gms02}%
  \BibitemOpen
  \bibfield  {author} {\bibinfo {author} {\bibfnamefont {D.~O.}\ \bibnamefont
  {Gericke}}, \bibinfo {author} {\bibfnamefont {M.~S.}\ \bibnamefont
  {Murillo}}, \ and\ \bibinfo {author} {\bibfnamefont {M.}~\bibnamefont
  {Schlanges}},\ }\href {\doibase 10.1103/PhysRevE.65.036418} {\bibfield
  {journal} {\bibinfo  {journal} {Phys. Rev. E}\ }\textbf {\bibinfo {volume}
  {65}},\ \bibinfo {pages} {036418} (\bibinfo {year} {2002})}\BibitemShut
  {NoStop}%
\bibitem [{\citenamefont {McQuillen}, \citenamefont {Castro},\ and\
  \citenamefont {Killian}(2011{\natexlab{a}})}]{mck11}%
  \BibitemOpen
  \bibfield  {author} {\bibinfo {author} {\bibfnamefont {P.}~\bibnamefont
  {McQuillen}}, \bibinfo {author} {\bibfnamefont {J.}~\bibnamefont {Castro}}, \
  and\ \bibinfo {author} {\bibfnamefont {T.~C.}\ \bibnamefont {Killian}},\
  }\href@noop {} {\bibfield  {journal} {\bibinfo  {journal} {J. Phys. B}\
  }\textbf {\bibinfo {volume} {44}},\ \bibinfo {pages} {184013} (\bibinfo
  {year} {2011}{\natexlab{a}})}\BibitemShut {NoStop}%
\bibitem [{\citenamefont {Killian}\ \emph {et~al.}(2012)\citenamefont
  {Killian}, \citenamefont {McQuillen}, \citenamefont {O'Neil},\ and\
  \citenamefont {Castro}}]{kmo12}%
  \BibitemOpen
  \bibfield  {author} {\bibinfo {author} {\bibfnamefont {T.~C.}\ \bibnamefont
  {Killian}}, \bibinfo {author} {\bibfnamefont {P.}~\bibnamefont {McQuillen}},
  \bibinfo {author} {\bibfnamefont {T.~M.}\ \bibnamefont {O'Neil}}, \ and\
  \bibinfo {author} {\bibfnamefont {J.}~\bibnamefont {Castro}},\ }\href
  {\doibase http://dx.doi.org/10.1063/1.3694654} {\bibfield  {journal}
  {\bibinfo  {journal} {Phys. Plasmas}\ }\textbf {\bibinfo {volume} {19}},\
  \bibinfo {eid} {055701} (\bibinfo {year} {2012})}\BibitemShut {NoStop}%
\bibitem [{\citenamefont {McQuillen}, \citenamefont {Castro},\ and\
  \citenamefont {Killian}(2011{\natexlab{b}})}]{pck11}%
  \BibitemOpen
  \bibfield  {author} {\bibinfo {author} {\bibfnamefont {P.}~\bibnamefont
  {McQuillen}}, \bibinfo {author} {\bibfnamefont {J.}~\bibnamefont {Castro}}, \
  and\ \bibinfo {author} {\bibfnamefont {T.~C.}\ \bibnamefont {Killian}},\
  }\href {http://stacks.iop.org/0953-4075/44/i=18/a=184013} {\bibfield
  {journal} {\bibinfo  {journal} {Journal of Physics B: Atomic, Molecular and
  Optical Physics}\ }\textbf {\bibinfo {volume} {44}},\ \bibinfo {pages}
  {184013} (\bibinfo {year} {2011}{\natexlab{b}})}\BibitemShut {NoStop}%
\bibitem [{\citenamefont {Gavrilyuk}, \citenamefont {Krasnov},\ and\
  \citenamefont {Shaparev}(1997)}]{gks97}%
  \BibitemOpen
  \bibfield  {author} {\bibinfo {author} {\bibfnamefont {A.}~\bibnamefont
  {Gavrilyuk}}, \bibinfo {author} {\bibfnamefont {I.}~\bibnamefont {Krasnov}},
  \ and\ \bibinfo {author} {\bibfnamefont {N.}~\bibnamefont {Shaparev}},\
  }\href {\doibase 10.1134/1.1261618} {\bibfield  {journal} {\bibinfo
  {journal} {Technical Physics Letters}\ }\textbf {\bibinfo {volume} {23}},\
  \bibinfo {pages} {61} (\bibinfo {year} {1997})}\BibitemShut {NoStop}%
\bibitem [{\citenamefont {Gavrilyuk}, \citenamefont {Krasnov},\ and\
  \citenamefont {Shaparev}(1998)}]{gks98}%
  \BibitemOpen
  \bibfield  {author} {\bibinfo {author} {\bibfnamefont {A.}~\bibnamefont
  {Gavrilyuk}}, \bibinfo {author} {\bibfnamefont {I.}~\bibnamefont {Krasnov}},
  \ and\ \bibinfo {author} {\bibfnamefont {N.~Y.}\ \bibnamefont {Shaparev}},\
  }\href@noop {} {\bibfield  {journal} {\bibinfo  {journal} {Laser Physics}\
  }\textbf {\bibinfo {volume} {8}},\ \bibinfo {pages} {653} (\bibinfo {year}
  {1998})}\BibitemShut {NoStop}%
\bibitem [{\citenamefont {Gavrilyuk}, \citenamefont {Krasnov},\ and\
  \citenamefont {Shaparev}(2002)}]{gks02}%
  \BibitemOpen
  \bibfield  {author} {\bibinfo {author} {\bibfnamefont {A.~P.}\ \bibnamefont
  {Gavrilyuk}}, \bibinfo {author} {\bibfnamefont {I.~V.}\ \bibnamefont
  {Krasnov}}, \ and\ \bibinfo {author} {\bibfnamefont {N.~Y.}\ \bibnamefont
  {Shaparev}},\ }\href@noop {} {\bibfield  {journal} {\bibinfo  {journal}
  {{JETP Lett.}}\ }\textbf {\bibinfo {volume} {76}},\ \bibinfo {pages} {423}
  (\bibinfo {year} {2002})}\BibitemShut {NoStop}%
\bibitem [{\citenamefont {Killian}\ \emph {et~al.}(2003)\citenamefont
  {Killian}, \citenamefont {Ashoka}, \citenamefont {Gupta}, \citenamefont
  {Laha}, \citenamefont {Nagel}, \citenamefont {Simien}, \citenamefont {Kulin},
  \citenamefont {Rolston},\ and\ \citenamefont {Bergeson}}]{kag03}%
  \BibitemOpen
  \bibfield  {author} {\bibinfo {author} {\bibfnamefont {T.~C.}\ \bibnamefont
  {Killian}}, \bibinfo {author} {\bibfnamefont {V.~S.}\ \bibnamefont {Ashoka}},
  \bibinfo {author} {\bibfnamefont {P.}~\bibnamefont {Gupta}}, \bibinfo
  {author} {\bibfnamefont {S.}~\bibnamefont {Laha}}, \bibinfo {author}
  {\bibfnamefont {S.~B.}\ \bibnamefont {Nagel}}, \bibinfo {author}
  {\bibfnamefont {C.~E.}\ \bibnamefont {Simien}}, \bibinfo {author}
  {\bibfnamefont {S.}~\bibnamefont {Kulin}}, \bibinfo {author} {\bibfnamefont
  {S.~L.}\ \bibnamefont {Rolston}}, \ and\ \bibinfo {author} {\bibfnamefont
  {S.~D.}\ \bibnamefont {Bergeson}},\ }\href@noop {} {\bibfield  {journal}
  {\bibinfo  {journal} {{ J. Phys. A: Math. Gen.}}\ }\textbf {\bibinfo {volume}
  {36}},\ \bibinfo {pages} {6077} (\bibinfo {year} {2003})}\BibitemShut
  {NoStop}%
\bibitem [{\citenamefont {Pohl}, \citenamefont {Pattard},\ and\ \citenamefont
  {Rost}(2004{\natexlab{c}})}]{ppr04prl}%
  \BibitemOpen
  \bibfield  {author} {\bibinfo {author} {\bibfnamefont {T.}~\bibnamefont
  {Pohl}}, \bibinfo {author} {\bibfnamefont {T.}~\bibnamefont {Pattard}}, \
  and\ \bibinfo {author} {\bibfnamefont {J.~M.}\ \bibnamefont {Rost}},\ }\href
  {\doibase 10.1103/PhysRevLett.92.155003} {\bibfield  {journal} {\bibinfo
  {journal} {Phys. Rev. Lett.}\ }\textbf {\bibinfo {volume} {92}},\ \bibinfo
  {pages} {155003} (\bibinfo {year} {2004}{\natexlab{c}})}\BibitemShut
  {NoStop}%
\bibitem [{\citenamefont {Krasnov}(2009)}]{kra09}%
  \BibitemOpen
  \bibfield  {author} {\bibinfo {author} {\bibfnamefont {I.}~\bibnamefont
  {Krasnov}},\ }\href {\doibase 10.1016/j.physleta.2009.04.042} {\bibfield
  {journal} {\bibinfo  {journal} {Physics Letters A}\ }\textbf {\bibinfo
  {volume} {373}},\ \bibinfo {pages} {2291} (\bibinfo {year}
  {2009})}\BibitemShut {NoStop}%
\bibitem [{\citenamefont {Gavriliuk}\ \emph {et~al.}(2009)\citenamefont
  {Gavriliuk}, \citenamefont {Isaev}, \citenamefont {Karpov}, \citenamefont
  {Krasnov},\ and\ \citenamefont {Shaparev}}]{gik09}%
  \BibitemOpen
  \bibfield  {author} {\bibinfo {author} {\bibfnamefont {A.}~\bibnamefont
  {Gavriliuk}}, \bibinfo {author} {\bibfnamefont {I.}~\bibnamefont {Isaev}},
  \bibinfo {author} {\bibfnamefont {S.}~\bibnamefont {Karpov}}, \bibinfo
  {author} {\bibfnamefont {I.}~\bibnamefont {Krasnov}}, \ and\ \bibinfo
  {author} {\bibfnamefont {N.}~\bibnamefont {Shaparev}},\ }\href {\doibase
  10.1103/PhysRevE.80.056404} {\bibfield  {journal} {\bibinfo  {journal} {Phys.
  Rev. E}\ }\textbf {\bibinfo {volume} {80}},\ \bibinfo {pages} {056404}
  (\bibinfo {year} {2009})}\BibitemShut {NoStop}%
\end{thebibliography}%

\end{document}